\documentclass[useAMS,usenatbib]{mn2e}
\usepackage{appendix}
\usepackage{color}
\usepackage{float}
\usepackage{graphicx}
\DeclareGraphicsExtensions{.pdf}
\usepackage{natbib}
\usepackage[normalem]{ulem}

\title[Gas and stellar populations in barred galaxies.]{Gaseous-phase metallicities and stellar populations in the centres of barred galaxies.}
\author[R. Cacho, et al.]{R. Cacho$^{1}$\thanks{E-mail:
raulcacho@ucm.es}, P. S\'anchez-Bl\'azquez$^{2}$, J. Gorgas$^{1}$, and I. P\'erez$^{3,4}$\\
$^{1}$Departamento de Astrof\'isica y Ciencias de la Atm\'osfera, Universidad Complutense de Madrid, E-28040, Madrid, Spain\\
$^{2}$Departamento de F\'isica Te\'orica, Universidad Aut\'onoma de Madrid, Cantoblanco, E-28049 Madrid, Spain\\
$^{3}$Departamento de F\'isica Te\'orica y del Cosmos, Universidad de Granada, E-18071, Granada, Spain\\
$^{4}$Instituto Universitario Carlos I de F\'isica Te\'orica y Computacional, E-18071, Granada, Spain}

\begin{document}

\date{Accepted 2014 May 07. Received 2014 April 21; in original form 2014 January 31}
\pagerange{\pageref{firstpage}--\pageref{lastpage}} \pubyear{0000}

\maketitle

\label{firstpage}

\begin{abstract} 
Numerical simulations predict that bars represent a very important agent for triggering gas inflows, which in turn could lead to central star formation. Bars thus are thought to contribute to the formation of the bulge.This changes both, the $\it{gaseous}$-phase and $\it{stellar}$-phase metallicities in the centres of galaxies. With the aim of quantifying the importance of this process we present a comparative study of the gaseous-phase and stellar-phase metallicities in the centres of members of a sample of barred and unbarred galaxies from SDSS. We do not find a significant difference in the metallicity (neither gaseous nor stellar) of barred and unbarred galaxies, but we find different trends in the metallicities of early- and late- type galaxies, with larger differences in the metallicity in the early-type subsample. Our results contradict some previous research in this field, but we find a possible origin of the discrepancies between previous works and our results.
\end{abstract}



\section{Introduction}\label{cap:introduction}

In the current model of cosmology, mergers are very important and drive galaxy evolution in the early stages of the Universe. At low redshift, however, the slow rearrangement of energy and mass that results from interactions involving collective phenomena, such as bars, oval discs, spiral structures and triaxial dark haloes, is believed to be at least as important as violent interactions \citep{Kormendy04, Beckman08}.

Numerical simulations of barred disc galaxies predict that bars are the most important internal mechanism for redistributing angular momentum in the baryonic and dark matter components \citep{Weinberg85, Debattista98, Debattista00, Athanassoula02, Berentzen06}. Bars also efficiently drive gas inflows into the central parts of the galaxy, feeding starburst, which might increase the central gaseous and stellar metallicity \citep{Elmegreen94, Knapen95, Hunt99, Jogee99, Jogee05, Jogee06}. This process can result in the formation of `discy' bulges. 

Several observational studies have tried to confirm these theoretical predictions by comparing the star formation rate (SFR) and the central metallicity (mainly from the ionized gas) of galaxies with and without bars. However, the results are far from being conclusive. For example, some studies have observed that barred galaxies have higher SFRs \citep{Hummel90,Martin95,Hawarden96,Huang96, Ellison11}, but these high SFRs do not apper to be explained by bars \citep{Pompea90, Martinet97, Chapelon99}. More discrepant results are found among studies of gaseous-phase metallicities: \cite{Henry99} found that barred and unbarred galaxies had the same gas metallicity at a given magnitude ($M_{B}$) while \cite{Dutil99} found that barred systems were more metal-poor (by 0.5~dex) at a given $M_{B}$ than unbarred and weakly barred galaxies. \cite{Considere00} also found lower central abundances for a given $M_{B}$ in their starbursting barred sample. However, with a larger sample of galaxies, Ellison et al. (2011, E11 hereafter) reported that metallicities of barred galaxies are uniformly higher than those of the unbarred sample by $\sim$ 0.06~dex at a given stellar mass. 

However, while a lot of attention has been paid to the gaseous-phase central metallicities, there is a dearth of studies analysing the stellar metallicites of galaxies as a function of bar presence. \cite{Moorthy06} found hints that, at a given velocity dispersion or rotational velocity, barred galaxies have higher central metallicities than unbarred galaxies, a result that was also found by \cite{Perez11}. However, in neither of these two works was the difference statistically significant. On the other hand \cite{Coelho12} found that central abundances of barred and unbarred galaxies were equivalent. However, they found significant differences ($\sim 4\sigma$) between the age distributions of the subsamples, although only for the most massive spirals.

It is worth noting that the present study is based on the idea that bars are long-lasting structures. There is no consensus in the literature about this point. Some authors argue that bars are easily destroyed by secular evolution \citep{Heller96, Berentzen98, Sellwood99, Shen04} although recent works, both theoretical and observational argue the contrary \citep{Berentzen07, Kraljic12, Athanassoula13}. If bars are not long-lasting structures, the fact that a galaxy does not have a bar does not mean that it did not have one in the recent past, and, therefore, could have been influenced by it.

Given the importance that bars may have for the evolution of disc galaxies, it is important to revisit this topic and to understand the sources of the discrepant results in previous works. We study the possible differences arising from sample selection, the measurement of fluxes and the methods used to calculate metallicities. We also present the first comparison of the stellar and gaseous abundances, as well as ages, for a sample of barred and unbarred galaxies covering the whole range of morphological types. The conclusions will be a key to understand the importance of secular processes in the building up of the central components of galaxies.

\section{THE SAMPLE}\label{cap:sample_selection}

We selected our sample from the catalogue of \cite{Nair10}. This catalogue contains more than 14.000 morphologically classified galaxies from SDSS with redshifts up to $0.1$ and brighter than 16 mag in the {\it g}-band. All galaxies in this catalogue were inspected visually (comparing $g$-band and $r$-band frames) to determine the presence and strength (considering strength as the ratio between the bar luminosity and galaxy luminosity) of a bar and were classified as strong, intermediate or weak\footnote{Note that even weak bars in this catalogue would be classified as strong bars in the RC3 catalogue \citep[see][for details]{Nair10}} \citep[with bar label 2, 4 or 8 in the][catalogue]{Nair10}. E11 performed an ellipse fitting to the galaxies from this catalogue, finding that the typical bar lengths of the barred sample is 3-10~kpc and the mean axial ratio $\sim$0.37. Our sample has the same characteristics.

We selected disc galaxies as those with morphological {\it T}-type $\geq-2$ (excluding all elliptical galaxies but including S0). We considered only pure bar galaxies; that is, those with bar label 2, 4 or 8. Unbarred galaxies are flagged with 0. Owing to the difficulty of observing bars in highly inclined galaxies and in order to minimize the disc contamination in the central spectrum, we limited our sample to those galaxies with a ratio between major and minor axes of $b/a\geq0.4$, which corresponds, roughly, to inclinations lower than $68^{\circ}$. However, we have checked that the results were {\bf not} affected by the inclination selection (at least when using $b/a\geq0.4$) by repeating the analysis using galaxies in different inclination bins (see Appendix~\ref{appendix_1}).

We obtained the line fluxes from the OSSY data base \citep{Oh11}. This catalogue includes all galaxies with $z\leq0.2$ and spectroscopic data from SDSS-DR7. The emission-line fluxes were measured using {\tt GANDALF} \citep[which also corrects for reddening by dust, see]{Sarzi06} combined with the MILES library \citep{Sanchez-Blazquez06, Vazdekis10}. {\tt GANDALF} fits, simultaneously, the stellar continuum and the emission-liness assuming Gaussians for the latter. This technique allows a superb subtraction of the stellar continuum that affects the Balmer line measurements \citep[for details see][]{Sarzi06}. In order to enable a comparison with the work by E11 we also took fluxes (not extinction-corrected) from the MPA/JHU\footnotemark[2]\footnotetext[2]{http://www.mpa-garching.mpg.de/SDSS/DR7/} data base. 

\begin{figure}
		\includegraphics[width=0.5\textwidth]{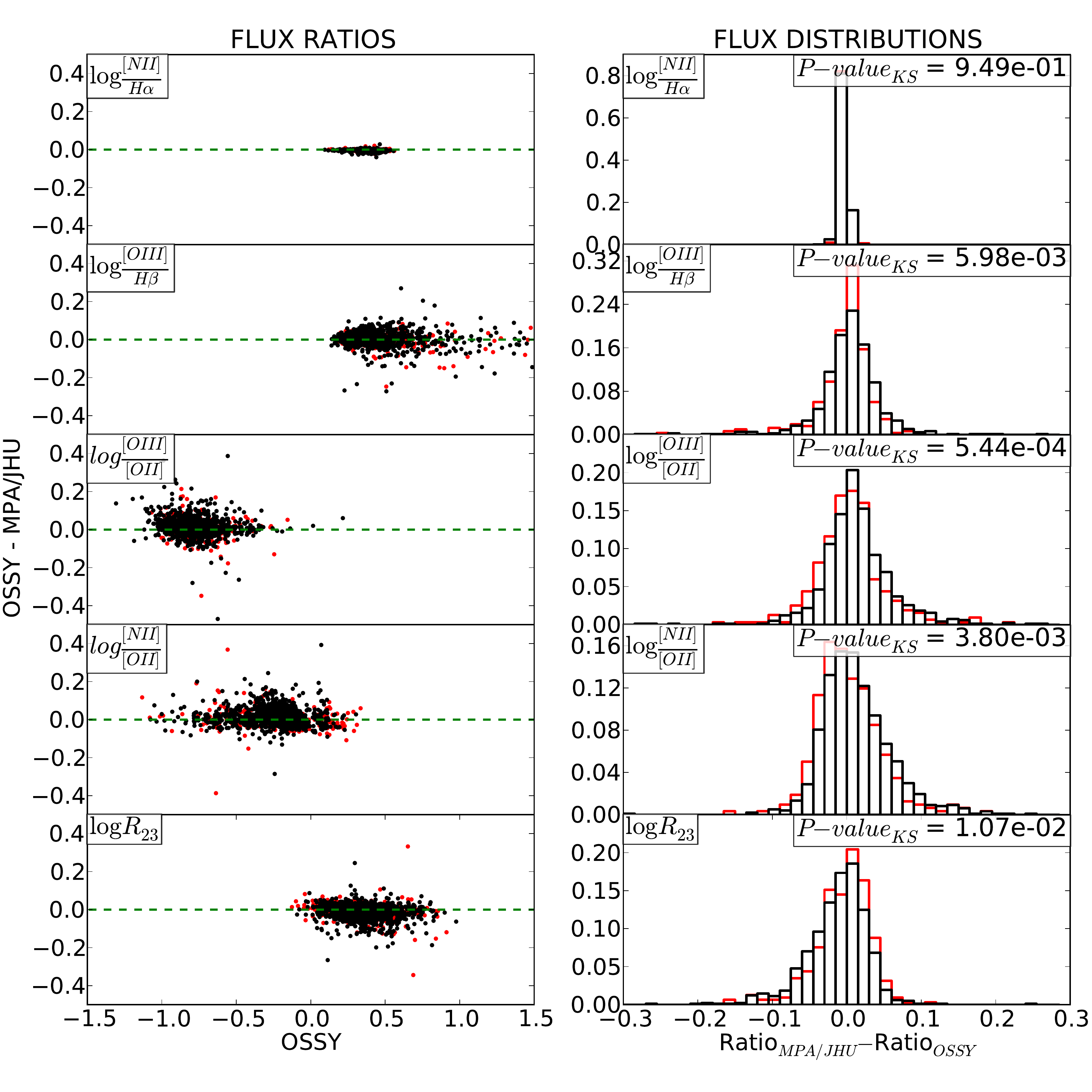}
 \caption{{\it Left panels:} Comparison of the flux ratios in different lines used in this work (from top to bottom: $\log \rm{[NII]/H}\alpha$, $\log \rm{[OIII]/H}\beta$, $\log \rm{[OIII]/[OII]}$, $\log \rm{[NII]/[OII]}$ and $\log \rm{R}_{23}$). {\it Right panels:} distributions of the flux ratios and results of KS tests. Red points correspond to barred galaxies, black points to unbarred galaxies and the green dashed line is 1:1. In the text boxes in the right panels we show the {\it P}-values for KS tests comparing the distributions of the flux ratios.}
 \label{figure_fluxes}
\end{figure}

Figure \ref{figure_fluxes} shows a comparison of the flux ratios used in different calibrations of the gas-phase metallicities between the two data bases. Flux ratios from OSSY and MPA/JHU correlate well with very low scatter (lower than that expected by the errors) except for $\log \frac{[OIII]}{[OII]}$, which is significantly larger (in mean) in barred galaxies by 0.01~dex. However, the flux measurements of the individual lines can differ considerably. Fluxes are correlated, but the scatter is large (and is larger for large fluxes) and the ratio $\frac{Flux_{OSSY}}{Flux_{MPA/JHU}}$ is far from 1:1, especially for oxygen lines (Florido et al. 2014, in prep.).

\begin{figure}
\includegraphics[width=0.5\textwidth]{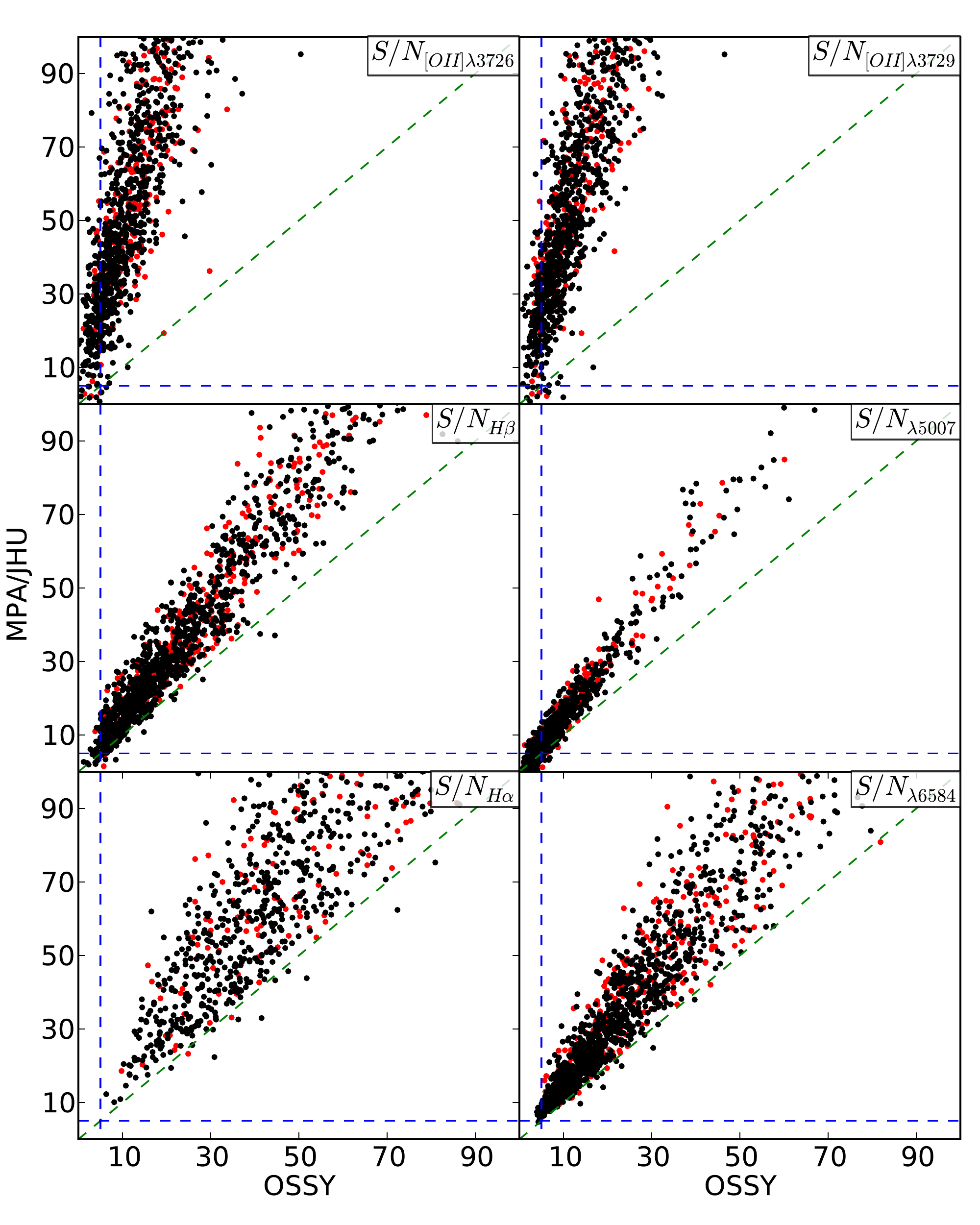}
\caption{Comparison of the $signal-to-noise$ ratios in OSSY \citep{Oh11} and MPA/JHU data bases. Red dots represent barred galaxies. black dots correspond to unbarred galaxies. The green dashed line correspond to 1:1. Blue dashed lines represent S/N=5 for each data base.}
\label{figureSN}
\end{figure}

We also compared the signal-to-noise ratios of the same emission-liness in the two data bases (see Fig. \ref{figureSN}). We expect a slope close to 1, but we observe a value close to 5/3. We will analyse later how the signal-to-noise affects the results.

\clearpage

We only selected galaxies if the strongest emission-liness ({\it i.e.}[OII]$\lambda3727$\footnotemark[1]\footnotetext[1]{This line is a doublet at $\lambda 3726$ and $\lambda 3729$ with the same flux each.}, H$\beta$, [OIII]$\lambda5007$, H$\alpha$ and [NII]$\lambda6584$) were detected with a minimum signal-to-noise, $S/N\geq3$ in order to avoid spurious detections. Owing to this criterion, applied on [OII]$\lambda3727$, and the spectral range of the SDSS spectrograph, a redshift cut appears at $z\geq0.02$.

\begin{figure}
\includegraphics[width=0.5\textwidth]{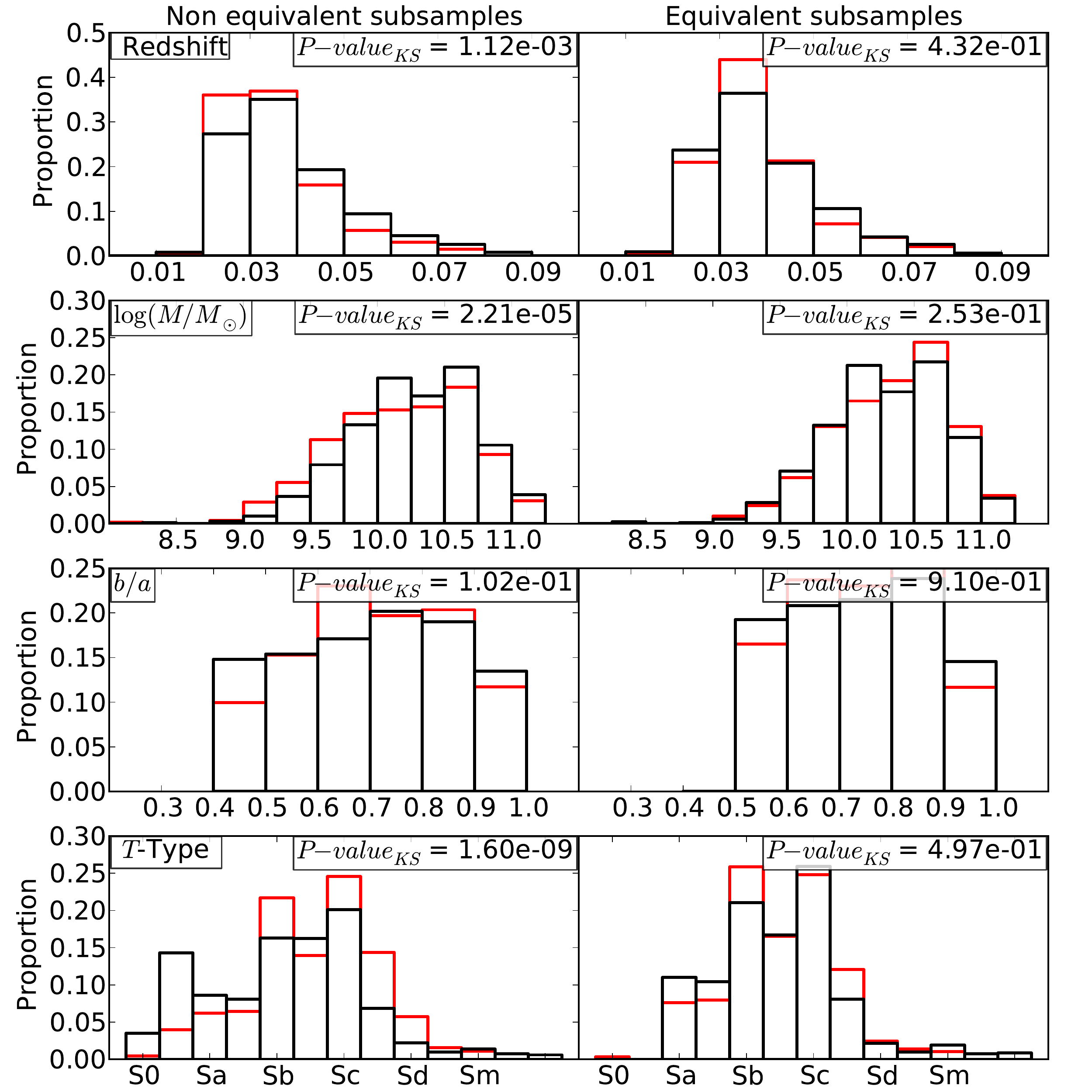}
\caption[width=1\textwidth]{{\it Left:} Distribution of the parameters in the non equivalent subsamples (see text for an explanation). {\it Right}: Distributions of the equivalent subsamples. Black and red lines represent the subsamples with $S/N\geq 3$. Overprinted are the {\it P}-values corresponding to the KS tests performed on the $S/N\geq3$ subsamples to check that barred and unbarred distributions are equivalent. These give the probabilities that the found differences are due to random effects.}
 \label{figureKSeq}
\end{figure}

Figure \ref{figureKSeq} shows the redshift, mass, $b/a$ and morphology distributions for our sample of barred and unbarred galaxies. All these parameters have been taken from the \cite{Nair10} catalogue.

The main goal of this study is to compare the central metallicities of barred and unbarred galaxies. As the differences between barred and unbarred galaxies may depend on their mass or their morphological type, as shown in various studies \citep[\textit{e.g.},]{Huang96,Ho97,James09, Coelho12}, we want to be sure that our subsamples are equivalent in those parameters. Furthermore, owing to possible biases with redshift and inclination, we also check if our subsamples of barred and unbarred galaxies share similar distributions in these parameters.

We use the mass as the reference in some sections in this paper. Mass is taken from the \cite{Nair10} catalogue, in which it is calculated using \cite{Kauffmann03a}. The mass is calculated using a synthetic library of star formation histories and the indices $D_{n}4000$ \citep{Balogh99} and $H\delta_{A}$ \citep{Worthey97}. With these indices, and by comparison with the SFH library, a mass-luminosity relation can be obtained for each individual galaxy, with an uncertainty of $\sim40$ per cent for a confidence level of 95 per cent.

We used the mass as the independent variable (instead of other direct measurements, such as luminosity) to enable comparison with previous studies that use mass. We checked, however, that our results are maintained using luminosity instead of mass as the independent variable.

A Kolmogorov-Smirnov (KS) test of the initial sample indicates that the two subsamples are not compatible with being taken from the same distribution. To remedy this, we removed random galaxies weighting in such a way that the probability of a galaxy being removed is higher if its properties are in those regions of the histogram where the distributions differ more. We did this until the distributions of all  four parameters considered, namely {\it M}, {\it z}, {\it b/a} and morphology, were equivalent for barred and unbarred galaxies (see Fig.~\ref{figureKSeq}).

After this process we had a sample of 414 barred galaxies and 1180 unbarred galaxies. The percentage of barred galaxies is 25.9 per cent, the same percentage as that in the catalogue of \cite{Nair10} (26 per cent).

\section{Analysis}
\subsection{Gaseous-phase metallicity}\label{cap:gas}

Before calculating metallicities, we identified and removed those galaxies in which the gas could be fully or partially ionized by an active galactic nucleus (AGN) using a BPT diagram \citep*{Baldwin81} with ${\rm [NII]}\lambda6584/{\rm H}\alpha$ and ${\rm [OIII]}\lambda5007/{\rm H}\beta$ (This diagram is shown in Fig.~\ref{figurebpt}). We used the separation criterion from \cite{Kauffmann03b}, which is based on empirical data and is more restrictive (we consider fewer galaxies as pure star-forming galaxies) than, for example, that of \cite{Kewley01}, which is based on photoionization models. After removing the galaxies with an AGN contribution, we calculated the metallicities using the calibrations described in \cite{Kewley02}. The method is an iterative process with the following steps:

\begin{figure}
\includegraphics[width=0.5\textwidth]{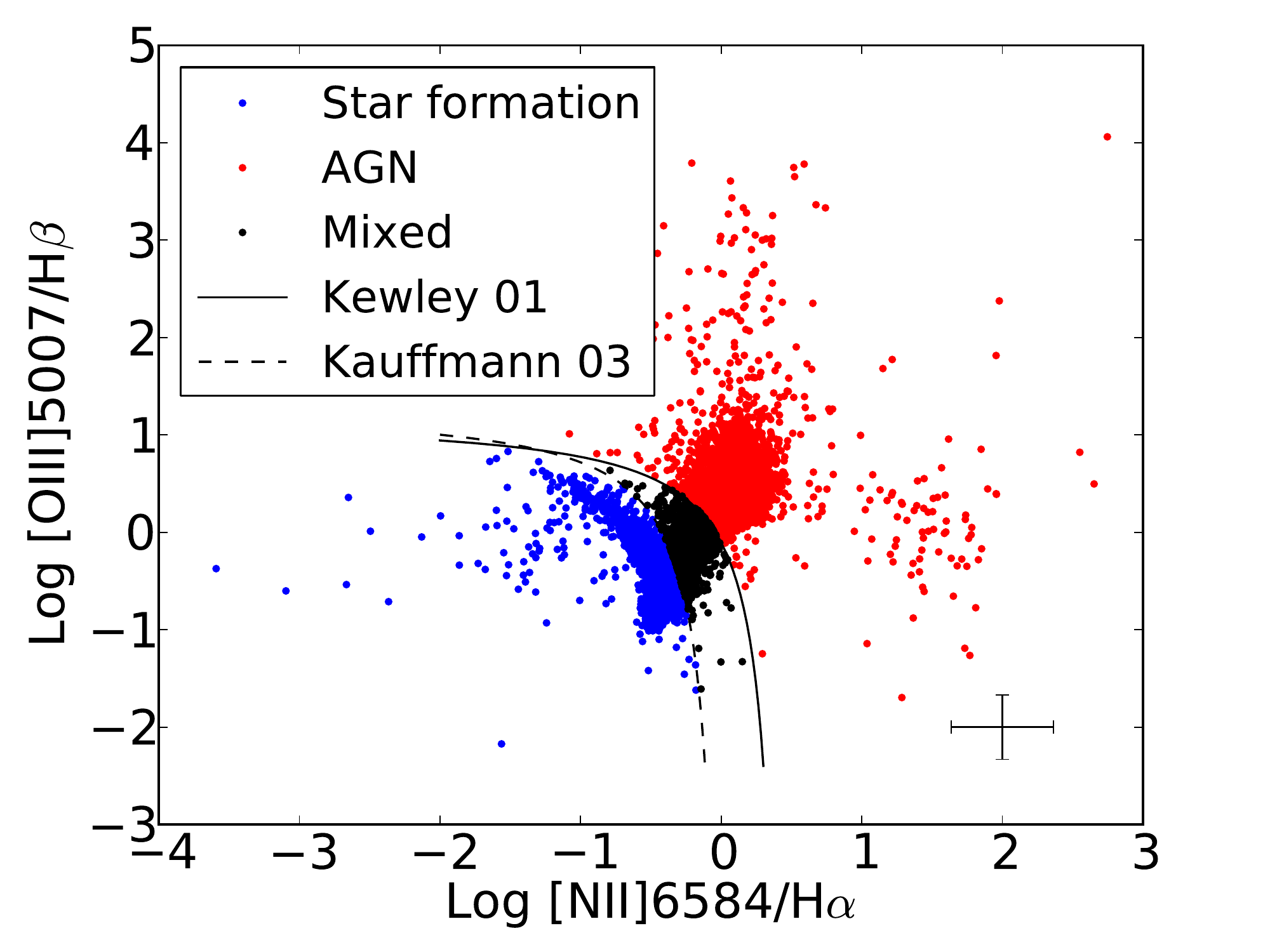}
\caption{BPT diagram. All galaxies are plotted here. The solid line represents the Kewley criterion and the dashed line the Kauffmann criterion. Red points are AGNs and LINERs, blue points are pure star forming galaxies and black points are star forming galaxies with some AGN contribution. Our sample comprises the galaxies in blue.}
\label{figurebpt}
\end{figure}

\begin{enumerate} 

\item Use ${\rm [NII]}\lambda6584/{\rm [OII]}\lambda 3727$ to derive a first guess for the metallicity value, which is almost independent of the ionization parameter. For low metallicities ($12 + \log({\rm O/H})<8.6$), the primary nitrogen is dominant and the calculated metallicities are not reliable.

\item Obtain a first value of the ionization parameter $q$, using the metallicity value from step (i). This is done using the ratio ${\rm [OIII]}\lambda5007/{\rm [OII]}\lambda3727$. 

\item Calculate a new value of the metallicity using the ionization parameter from step (ii), the coefficient $R_{23}$ and the polynomials in \cite{Kewley02}. The metallicity obtained in the first step is used to distinguish between high- and low- metallicity regimes, as $R_{23}$ is bi-valued.

\item Use this new value of metallicity and repeat this process from (ii) until both $q$ and $12 + \log({\rm O/H})$ converge to a precision of 0.01~dex. This typically happened in one or two iterations.

\end{enumerate}

We ran this process 100 times for each galaxy to derive uncertainties. Each time, we introduced random values for ${\rm [NII]}/{\rm [OII]}$, ${\rm [OIII]}/{\rm [OII]}$ and $R_{23}$, distributed following a Gaussian curve centred on the measured value and with a width equal to the uncertainty of each parameter. This method caused some calculations to fail because the mean value of $R_{23}$ was near the maximum. We computed the final metallicity as the mean of the output values and the associated uncertainty as its root mean square (RMS) dispersion.

Note that, despite E11 use the same iterative method than us, they use the calibration of ([NII]/[OII]) instead of that of $R_{23}$ to obtain metallicities. We use $R_{23}$ as metallicity calibration because it takes into account the ionization parameter, which is different for barred and unbarred galaxies, as will be explained in Sec. \ref{cap:star_formation_rate}. However, we compared all our results obtained with $R_{23}$ with the results obtained with [NII]/[OII] (see Sec. \ref{Previous_gas}).

\subsection{Stellar population parameters}

We characterized the stellar populations properties using the code {\tt STECKMAP} \citep{Ocvirk06a, Ocvirk06b}, which is a method for recovering the kinematic properties and the stellar content from the integrated spectra of galaxies. The reconstruction of the stellar age, the age-metallicity relation and the line-of-sight velocity distribution with this method are all non-parametric; that is, no specific shape is assumed. The only a priori conditions are the positivity and the requirement that the solution is smooth enough \citep[for details see][]{Ocvirk06a, Ocvirk06b, Sanchez-Blazquez11}.

From the recovered star formation history (SFH) and the age-metallicity relation, we obtain mean values of age and metallicity by weighting either in mass or in light. Here we use luminosity-weighted values because these are more directly associated with observations and are more robustly derived \citep[see, {\it e.g.},][]{Sanchez-Blazquez11}. In order to asses the robustness of our results to the method we also ran the code \citep[{\tt STARLIGHT},][]{CidFernandes05}. In Appendix \ref{Starlight_Steckmap_Lick} we show this comparison. As can be seen, while there are strong differences in the mass-weighted values derived with the two codes, the luminosity-weighted ages and metallicities are very similar independently of the code used to derive them. Luminosity-weighted values are more robust than mass-weighted ones, as the contribution of old, low-mass, low-luminosity stars can change considerably without changing the observed spectra much.

We also compared the values derived with {\tt STECKMAP} with those derived directly from Lick/IDS indices \citep{Worthey94} by comparing the positions of the galaxies in a ${\rm H}\beta$ versus {\rm Fe4668} diagram with the predictions of single-stellar population (SSP) models (also shown in Appendix \ref{Starlight_Steckmap_Lick}). These two values are not exactly comparable (as the values obtained with the indices are SSP-equivalent parameters, and do not necessarily coincide with the luminosity-weighted ones). However, despite the slope in the comparison of the two values differing from the 1:1 relation, the correlation is evident and the scatter is compatible with the measurement errors (see right panel in Fig. \ref{figure_indicessynthesis} in Appendix \ref{Starlight_Steckmap_Lick}).

\section{Results}
\label{gas_results} 

\subsection{Differences in the mass-metallicity relation}
\label{gas_mass_metallicity}

\begin{figure}
\includegraphics[width=0.5\textwidth]{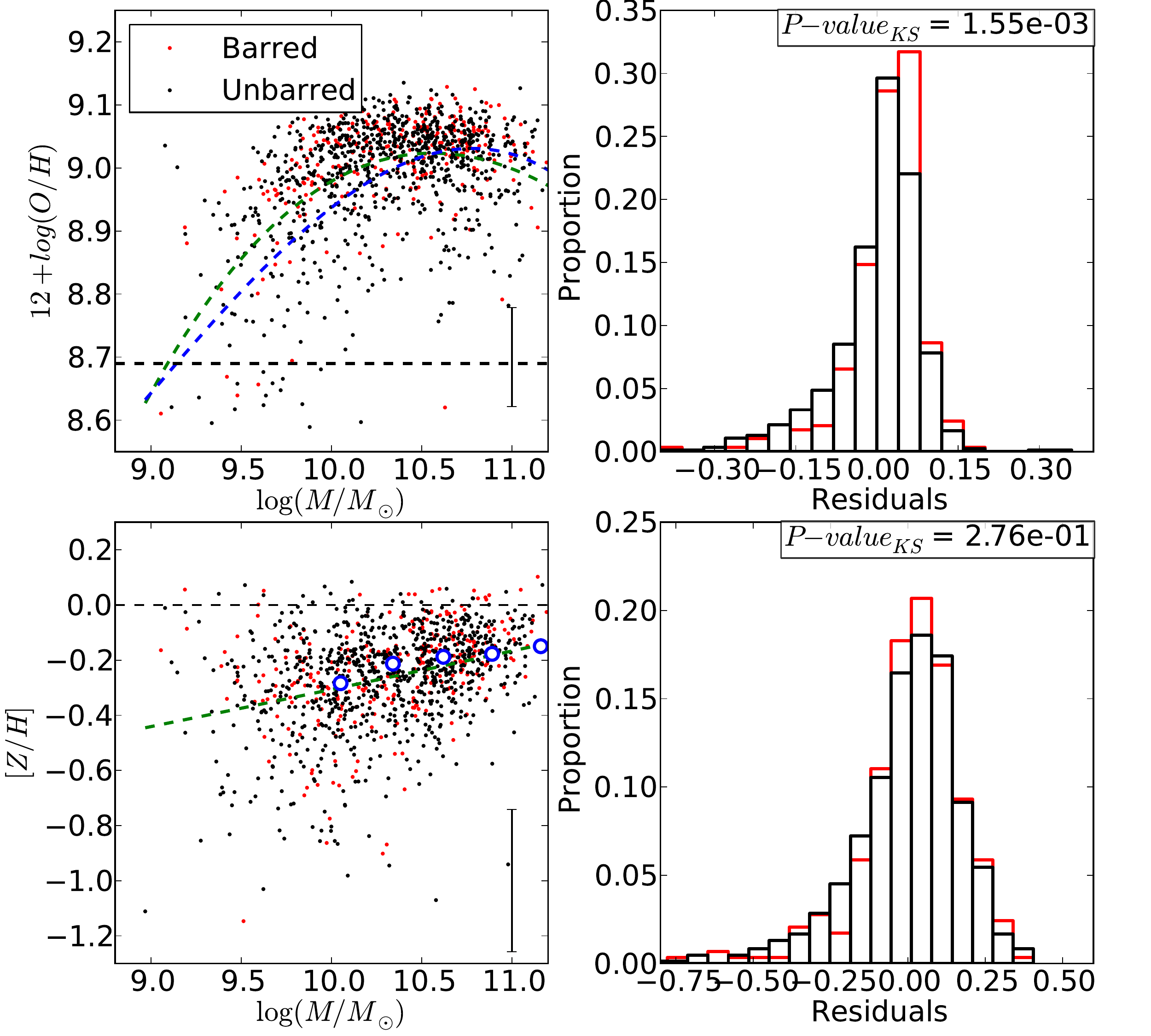}
\caption{Mass–metallicity relations. (Upper panel: left) gaseous metallicity
versus the stellar mass for all galaxies (red for barred, black for unbarred).
Green and blue lines show the polynomial fit to our data and that found by
\citealp{Kewley08}, respectively. Error bars represent the typical uncertainty
in the metallicities. (Lower panel: left) As upper panel, but for the stellar metallicity. Green dashed line represents a polynomial fit to the data. Blue open circles in the lower left panel represent the mass–metallicity relation found by \citealp{ValeAsari09}, shifted down by 0.25 dex. (Right panels) Distributions of the residuals from the polynomial. The text in boxes shows {\it P}-values for the KS tests (i.e. probabilities that the found differences are random effects).}
\label{figureMassMetal}
\end{figure}

Figure~\ref{figureMassMetal} shows both the stellar and gaseous-phase metallicities versus the stellar mass for our sample of galaxies. To check for possible differences in the central metallicity between barred and unbarred galaxies at a given stellar mass, we compare the residuals from a fitted third-order polynomial ($-9.92 + 4.15 x -0.28 x^2 + 0.0051 x^3$, with $x=\log M_{*}$) to the whole sample. We also plot, in Fig.~\ref{figureMassMetal}, the relation obtained by \cite{Kewley08}. We compare the residuals from the fit for both, barred and unbarred galaxies in the same figure.

If we average all the differences for the two subsamples, we find that the mean gaseous-phase metallicity of all barred galaxies is, on average, 0.02~dex higher than the mean metallicity of all unbarred galaxies. A KS test performed on the distribution confirms that the differences are statistically significant (the probability that the measured differences are obtained by chance is {\it P}-value = 8.8$\times 10^{-5}$).

\begin{figure}
\includegraphics[width=0.5\textwidth]{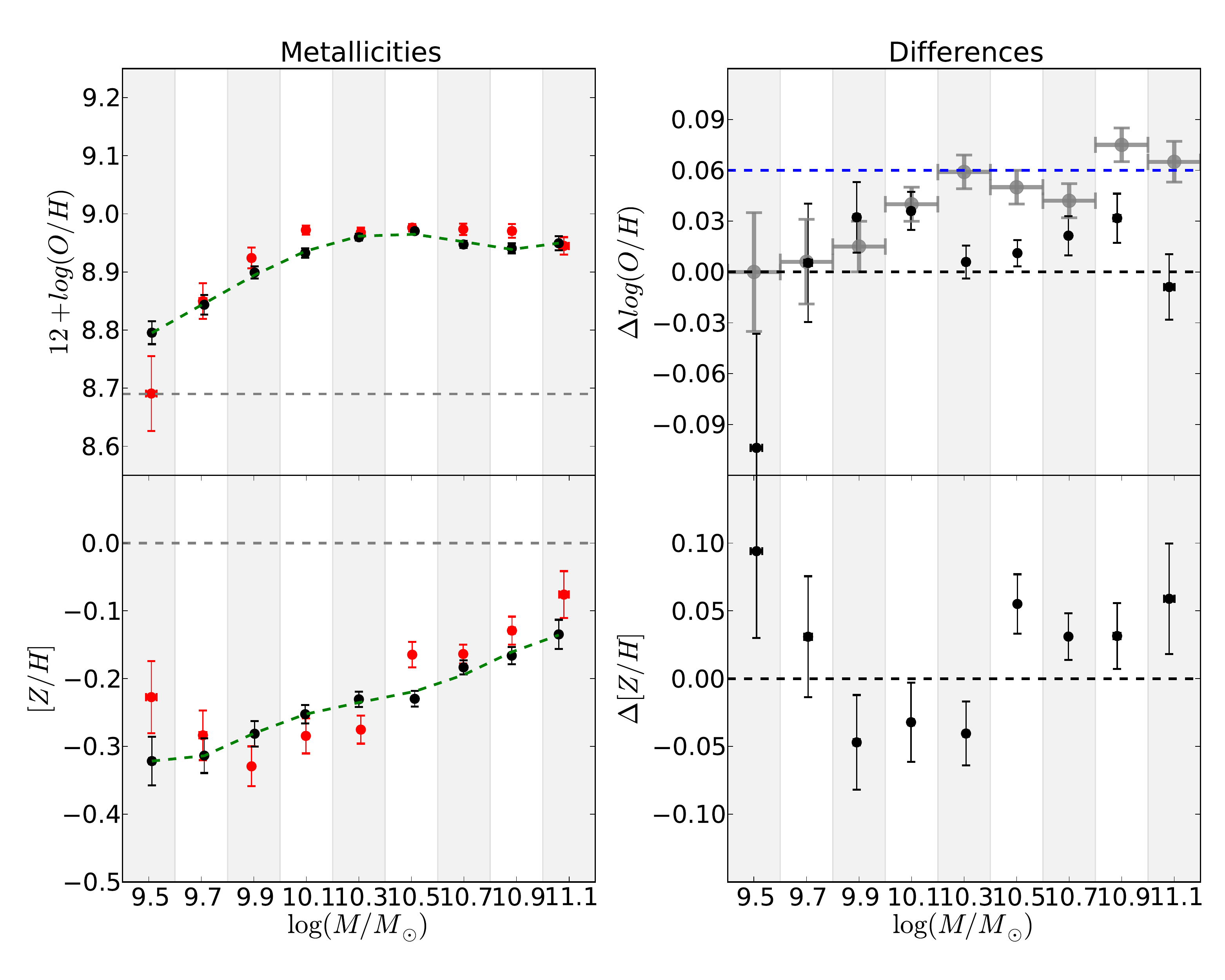}
\caption{(Upper panel: left) Mean metallicities versus the mean mass in each interval for barred galaxies (red points) and unbarred galaxies (black points) and the polynomial fit to unbarred galaxies (dashed green line). (Right) Residuals of barred galaxies from the polynomial. Black points represents not significant differences at 3 sigma level (corresponding to $P$=0.0027). Round grey points represent the differences found by \citealp{Ellison11}. Error bars represent the error in the mean. (Lower panels) Results for the stellar metallicities.}
 \label{figure_binmass}
\end{figure}

However, the mean differences might depend on the galaxy mass. To check this point we binned the galaxies in mass intervals. We first checked that the subsamples of barred and unbarred galaxies in each mass bin were equivalent in terms of redshift, inclination ($b/a$) and {\it T}-type distributions (we checked this with KS tests in each interval). We calculated the mean metallicity and mass of the galaxies in each bin for the gaseous and stellar phases. Figure \ref{figure_binmass} (upper panel) shows the residuals of barred galaxies from a polynomial fit to the points corresponding to unbarred galaxies.

As can be seen, in each individual mass bin, the differences between barred and unbarred galaxies are very small and, in fact, they are not significant in any case. However, the differences show a systematic behaviour, with barred galaxies having higher metallicities. The differences calculated here are always smaller than those found in E11. The mean difference, considering all binnings with more than five galaxies, is 0.019$\pm$0.002~dex (compared to 0.06~dex found by E11) which is compatible with the differences found globally (0.015$\pm$0.006~dex). However, we stress that, in each mass bin, the differences are not statistically significant.

We repeated the process for the stellar metallicities, $[Z/H]$, using the luminosity weighted values obtained with {\tt STECKMAP}. Figure \ref{figureMassMetal} shows the results. In this case, the KS test shows that the mean difference between barred and unbarred galaxies (0.015~dex) is not significant.

We compared the Mass-Metallicity (M-Z) relation we found with the relation found by \cite{ValeAsari09}. Despite different methodologies ({\tt STECKMAP} versus {\tt STARLIGHT}), sample selection (SDSS-DR4 versus SDSS-DR5), different limiting magnitudes, redshift range, S/N cuts, etc.), and different SSPs, we found an offset of only 0.25 dex between their M-Z relation and ours. 

Figure \ref{figure_binmass} (lower panel) shows the differences for the stellar-phase metallicities divided in mass bins. On the left we plot the mean masses and metallicities of barred and unbarred galaxies in each mass interval and a polynomial fit to the unbarred galaxies. On the right, we show the differences in the metallicities between barred galaxies and the polynomial. The differences in abundances are compatible with the error bars, with $P$-values$\geq$0.045 (at a 3$\sigma$ level and the sizes of our subsamples used, we would not be able to detect differences smaller than 0.04~dex, in mean), so the differences in the stellar metallicity of barred and unbarred galaxies at a given mass are not significant.

\subsection{Morphology}\label{cap:section_gas_morphology}

Some authors have found that the differences in the properties of barred and unbarred galaxies are visible only in early-type spirals \citep{Ho97, Coelho12}. This is believed to be because these galaxies host stronger bars \citep{Erwin05}. In order to check a possible dependence of our results on morphological type, we divided our sample of galaxies into two groups; the first group contains those galaxies with morphological types from S0 to Sb ({\it i.e.} $-2\leq$ {\it T}-type$\leq 3$); the second group includes galaxies from Sb to Sm ({\it i.e.} {\it T}-type$>$3). 

\begin{figure}
\includegraphics[width=0.5\textwidth]{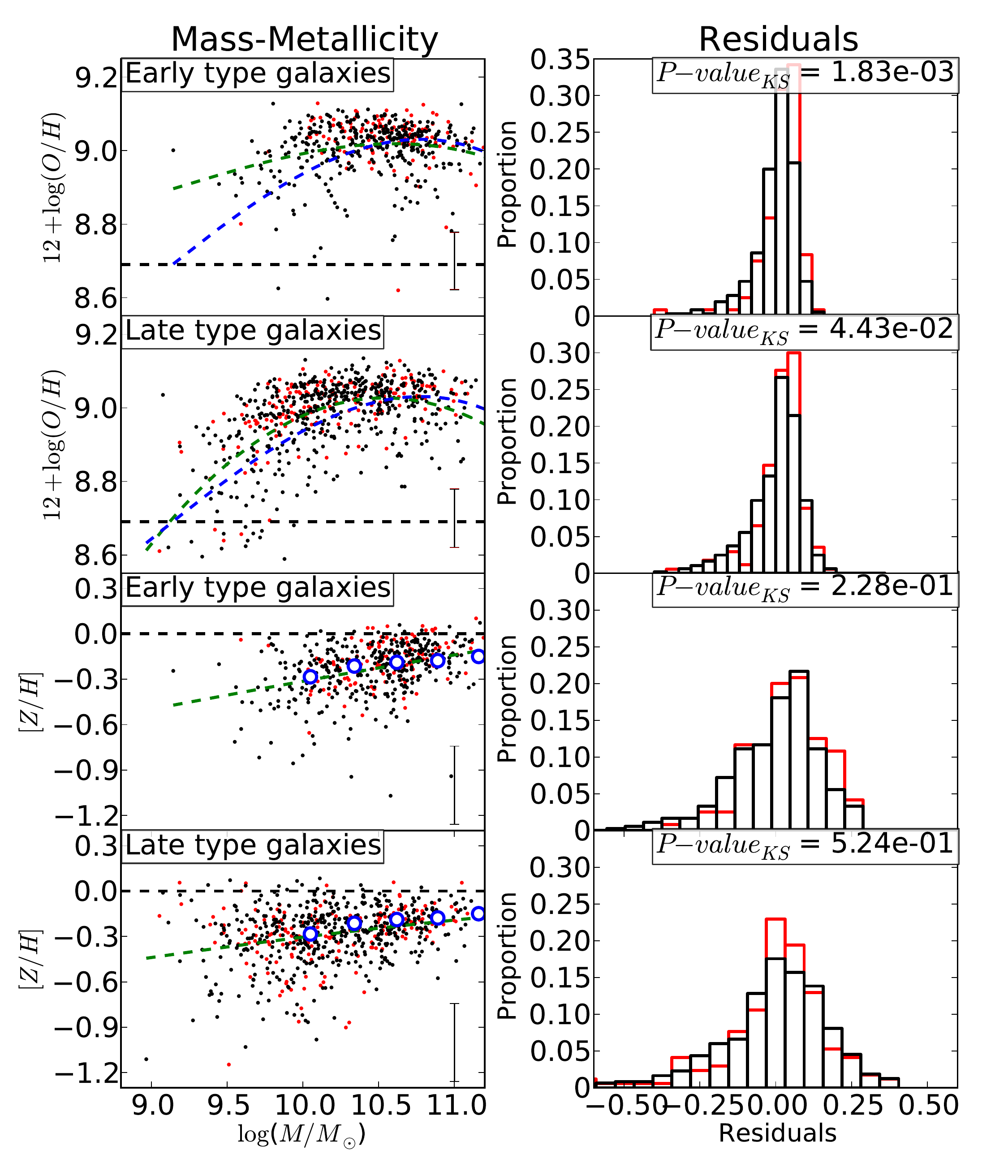}
\caption{Gaseous-phase (upper two panels) and stellar-phase (bottom two panels) metallicities
 versus mass for early ({\it T}-type $<$3) and late ({\it T}-type $>3$) disc galaxies. Left panels show the metallicity versus the mass for barred galaxies (red points) and unbarred galaxies (black points) and a polynomial fit (dashed line) to the whole sample of galaxies. Green and blue lines represent the same as in Fig. \ref{figureMassMetal}. Right panels show the residuals from the polynomial fit. Red lines represent barred galaxies and black lines represent unbarred galaxies. Text boxes on right panel show the $P$-values for KS tests (probability that the differences come from random effects).}
\label{figure_morpho}
\end{figure}

Figure~\ref{figure_morpho} shows a comparison between the metallicities of the two subsamples. For the gaseous-phase metallicity, we find mean differences in the metallicity of barred and unbarred galaxies of 0.027 $\pm$ 0.007~dex and 0.010 $\pm$ 0.008 for early- and late-type galaxies respectively. That is, we find stronger differences in the early-type sample.

\begin{figure}
\includegraphics[width=0.5\textwidth]{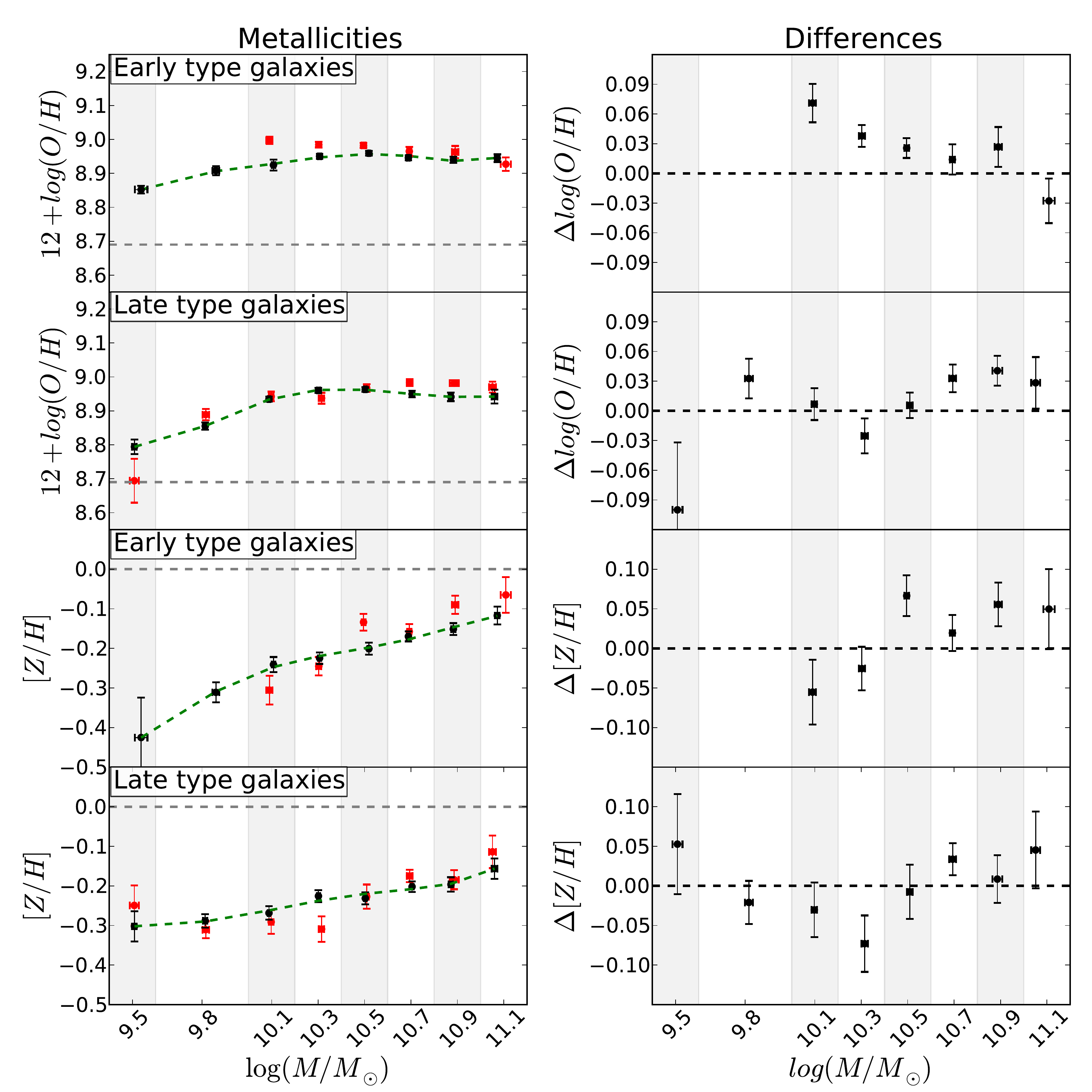}
\caption{As Fig. \ref{figure_binmass} but referred to early (first and third rows) and late (second and fourth rows) type galaxies. There are no significant differences ($P<$ 0.027) between barred and unbarred galaxies.}
\label{figure_morphobin}
\end{figure}

In Fig. \ref{figure_morphobin} the subsamples are binned in mass intervals. Barred galaxies are systematically more metal-rich than unbarred galaxies, which is more evident in the early-type group. These differences are not significant in any of the bins, but different trends can be seen, with the differences in the early-type group becoming smaller for increasing masses, and a flatter trend with the mass for the late-type galaxies. By comparing the number of galaxies in each bin, we found that the stellar mass distribution is different in barred and unbarred galaxies. If the sample is biased to early type galaxies, there could be differences in the metallicity owing to the different mass distributions.

\subsection{Bar strength}\label{cap:section_bar_strength}
Numerical simulations indicate that only strong bars are effective at funnelling gas to the inner kiloparsecs \citep[\textit{e.g.}][]{Regan04}. In order to check whether the central metallicities depend on the strength of the bar, we repeated the analysis after separating the galaxies in our sample in four subsamples: unbarred galaxies, weak bar galaxies, medium bar galaxies and strong bar galaxies attending to the `Bar' label in the \cite{Nair10} catalogue. This parameter is used as a proxy for the bar strength. 

\begin{figure*}
\includegraphics[width=1\textwidth]{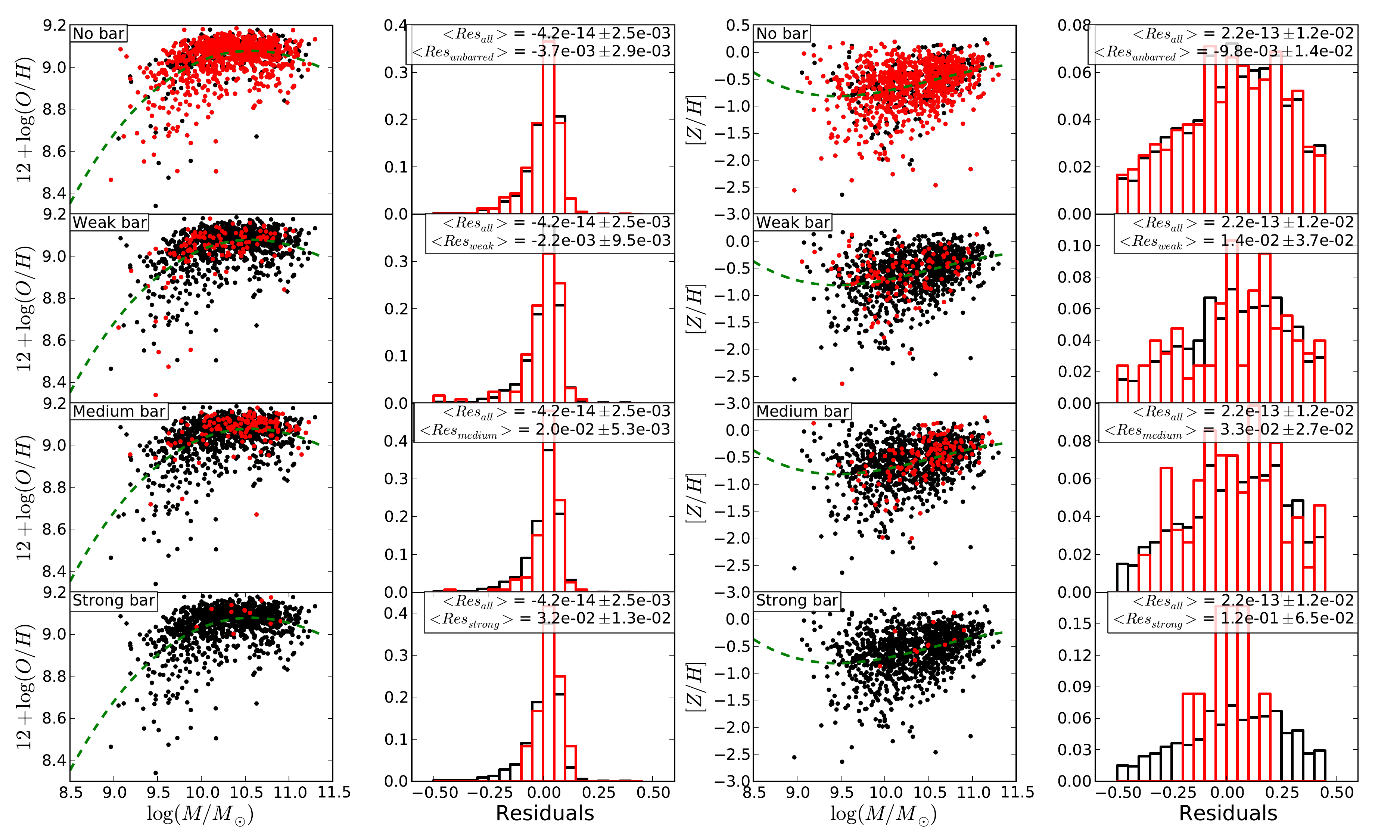}
\caption{Stellar and gaseous metallicities versus the stellar mass for unbarred (first row), weak bar (second row), medium bar (third row) and strong bar (fourth row) galaxies. In these plots, black points represent the whole sample of galaxies and red points the subsample described on the top lef corner in the panels. Black (red) lines represent the histogram of the residuals of black (red) points from the third-order polynomial fit (green dashed line) to the black points. On the top right corners there are overprinted the mean residuals of the whole sample and the subsample. None of them are significantly different from zero, in accordance with the T-test ($P$-value$\geq 0.005$).}
 \label{figure_bar_strength}
\end{figure*}

Fig. \ref{figure_bar_strength} shows the (gaseous and stellar) metallicities versus the mass for these subsamples compared to the whole sample. We calculated the residuals of the subsamples from a third-order polynomial fit to the whole sample and compared the residuals with the residuals of the whole subsample. As can be seen the mean of the residual increases from unbarred to strongly barred galaxies but it is never significantly different from zero.

\subsection{Star formation rate}\label{cap:star_formation_rate}

\begin{figure}
\includegraphics[width=0.5\textwidth]{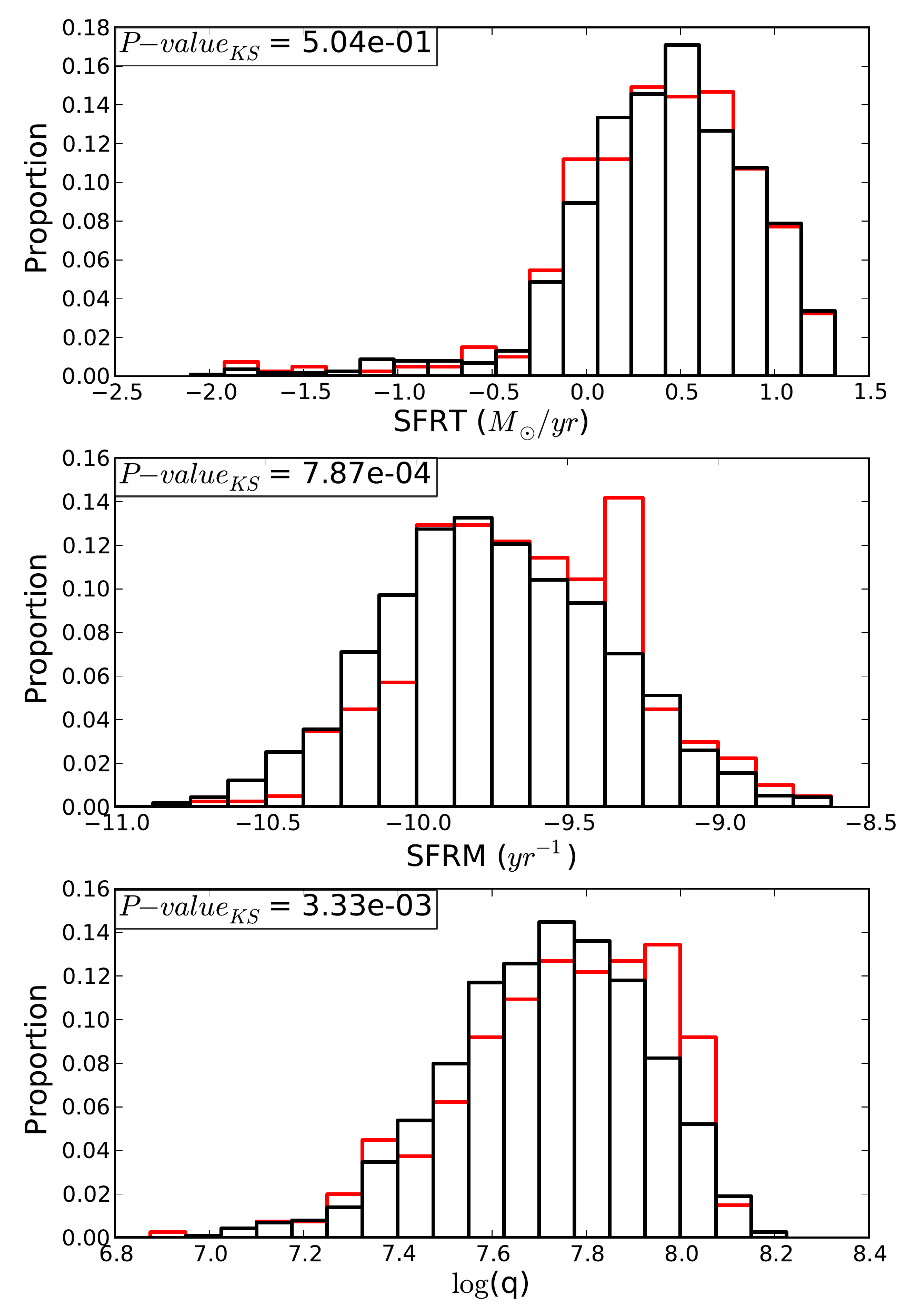}
\caption{From top to bottom: total SFRs, SFRs per unit mass and ionization parameters for barred (red) and unbarred (black) galaxies. The text boxes give the probability that the differences are random effects.}
\label{figure_SFR_logq}
\end{figure}

Figure \ref{figure_SFR_logq} shows a comparison between the SFRs and the ionization parameter between barred and unbarred galaxies. Barred and unbarred galaxies have the same SFR at 3$\sigma$ (with $P$-value=0.06), but the specific SFR is higher in barred galaxies at the same confidence level owing to the enhancement of the SFR in barred galaxies \citep[][]{Hawarden86,Ho97}. The ionization parameter is also higher in barred galaxies;that is, the number of massive stars (ionizing sources) is higher in barred galaxies.

As explained in Section \ref{cap:sample_selection}, the ratio [OIII]/[OII] is significantly different in barred and unbarred galaxies. This ratio affects directly to the calculation of the ionization parameter ($\log(q)$), which is involved in the calculation of the metallicity. The [NII]/[OII] is almost insensitive to $\log(q)$, while this is not true for $R_{23}$. A barred and an unbarred galaxy with the same [NII]/[OII], $R_{23}$, but different [OIII]/[OII] ratios would have the same metallicity if [NII]/[OII] is used, but there are differences ($\sim$ 0.3 dex) if $R_{23}$ is chosen as calibration.


\section{Gas versus stellar parameters}\label{cap:gaseous_vs_stellar}

If gaseous and stellar metallicities are linked through the SFH, a relationship between them should exist. 
A comparison of these two quantities can help to clarify whether bars trigger episodes of star formation in the centres of galaxies.

\begin{figure}
\includegraphics[width=0.5\textwidth]{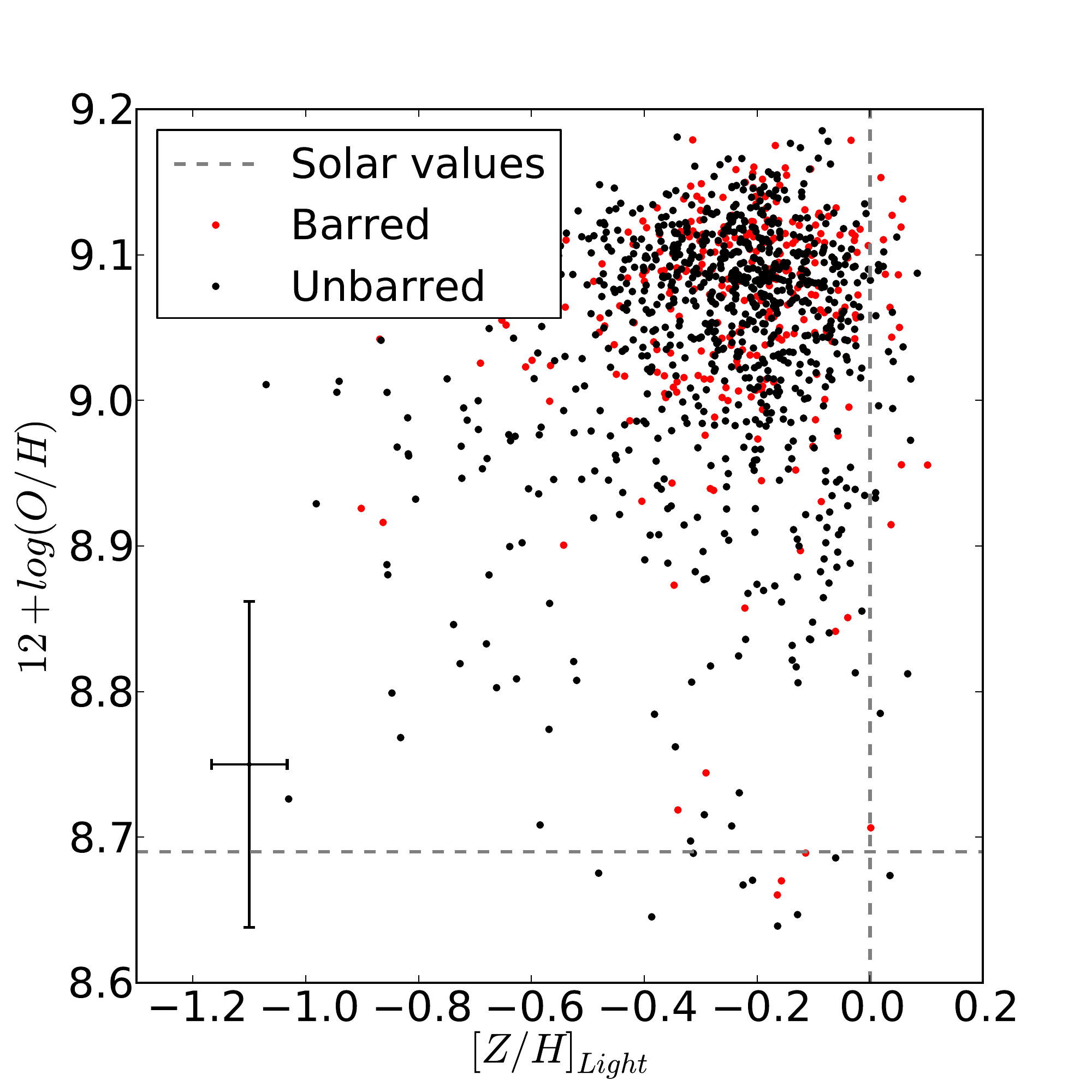}
\caption{Comparison of the stellar and gaseous metallicities. The error bars represent the typical errors for each galaxy. The uncertainties in the gaseous metallicities are affected by an error in the calibration of 0.07~dex.}
\label{figure_metallicities}
\end{figure}

Fig.~\ref{figure_metallicities} shows $12+\log ({\rm O}/{\rm H})$ versus $Z$. As can be seen, there is no correlation between gaseous and stellar metallicities (Spearman coefficient are $\rho _{barred}$=0.10 and $\rho _{unbarred}$=0.04 with $P$-values 0.069 and 0.17, respectively). We also compared the gaseous-phase metallicity versus Mgb$\lambda 5577$ index, but no correlation was observed.

\begin{figure}
\vspace{0pt}
\includegraphics[width=0.5\textwidth]{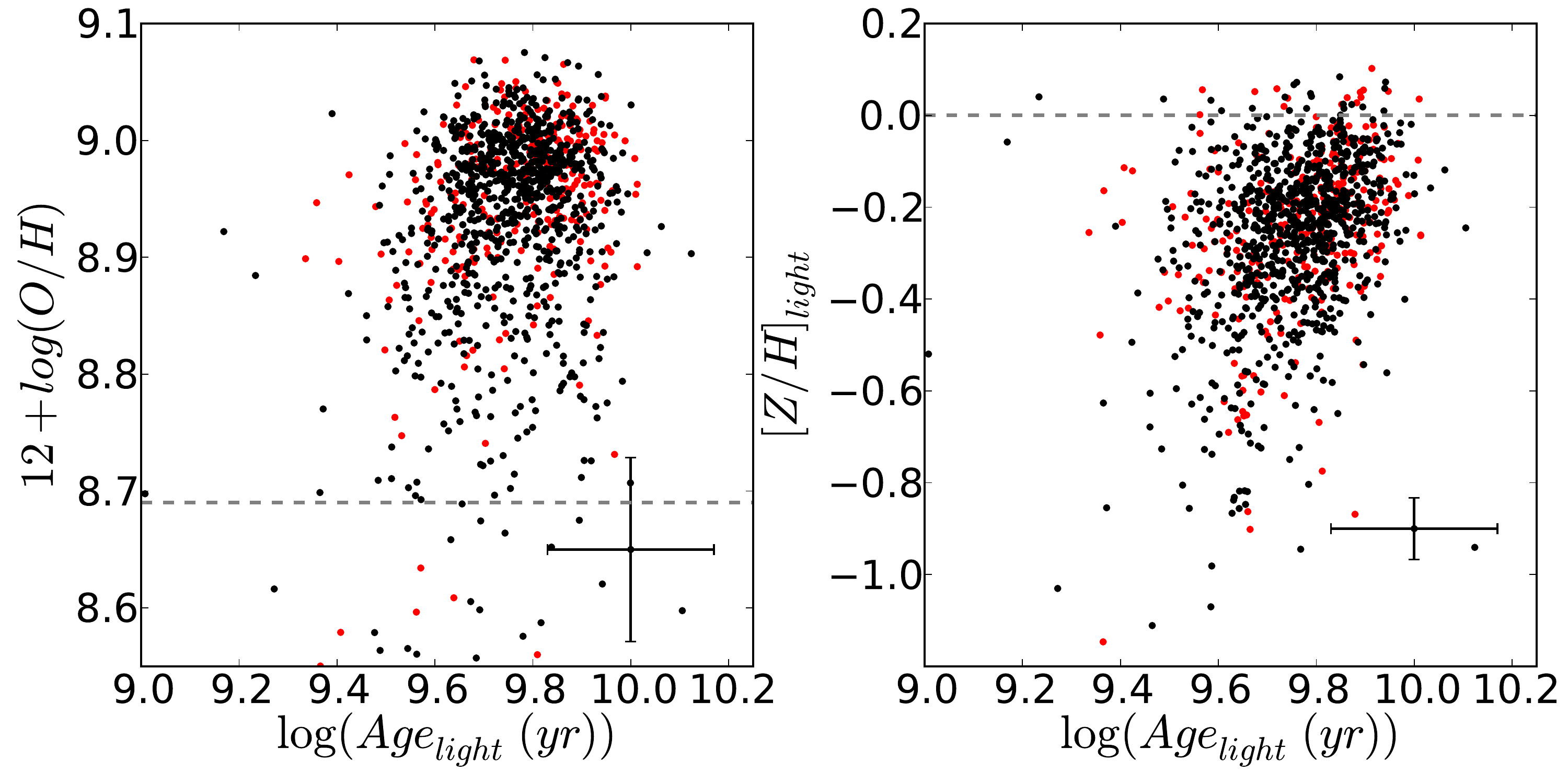}
\caption{Plots of the gaseous (left panel) and stellar (right panel) metallicities versus the stellar age. Red points represent barred galaxies, black points represent unbarred galaxies and dashed grey lines represent solar values. The error bars represent the typical uncertainties on the corresponding quantities.}
\label{figure_metalages}
\end{figure}

Figure \ref{figure_metalages} shows the gaseous and stellar metallicities versus the stellar age. A higher-limit in the metallicity can be seen for both gaseous (12+$\log$(O/H)$\sim$9.1) and stellar ([Z/H]$\sim$0.1) phases.

The right panel in Fig. \ref{figure_metalages} shows that older galaxies are also more metal-rich. The Spearman correlation coefficients are $\rho_{barred}$=0.39 $\rho_{unbarred}$=0.36 with $P$-values 5$\times 10^{-13}$ and 3$\times 10^{-32}$. Note that the age-metallicity degeneracy produces an artificial anti-correlation between these two parameters. Therefore, the fact that we obtain a positive correlation here means that full spectral fitting is a better tool than the classical index-index diagram to break the age-metallicity degeneracy \citep[see][]{Sanchez-Blazquez11}. 

\begin{figure}
\includegraphics[width=0.5\textwidth]{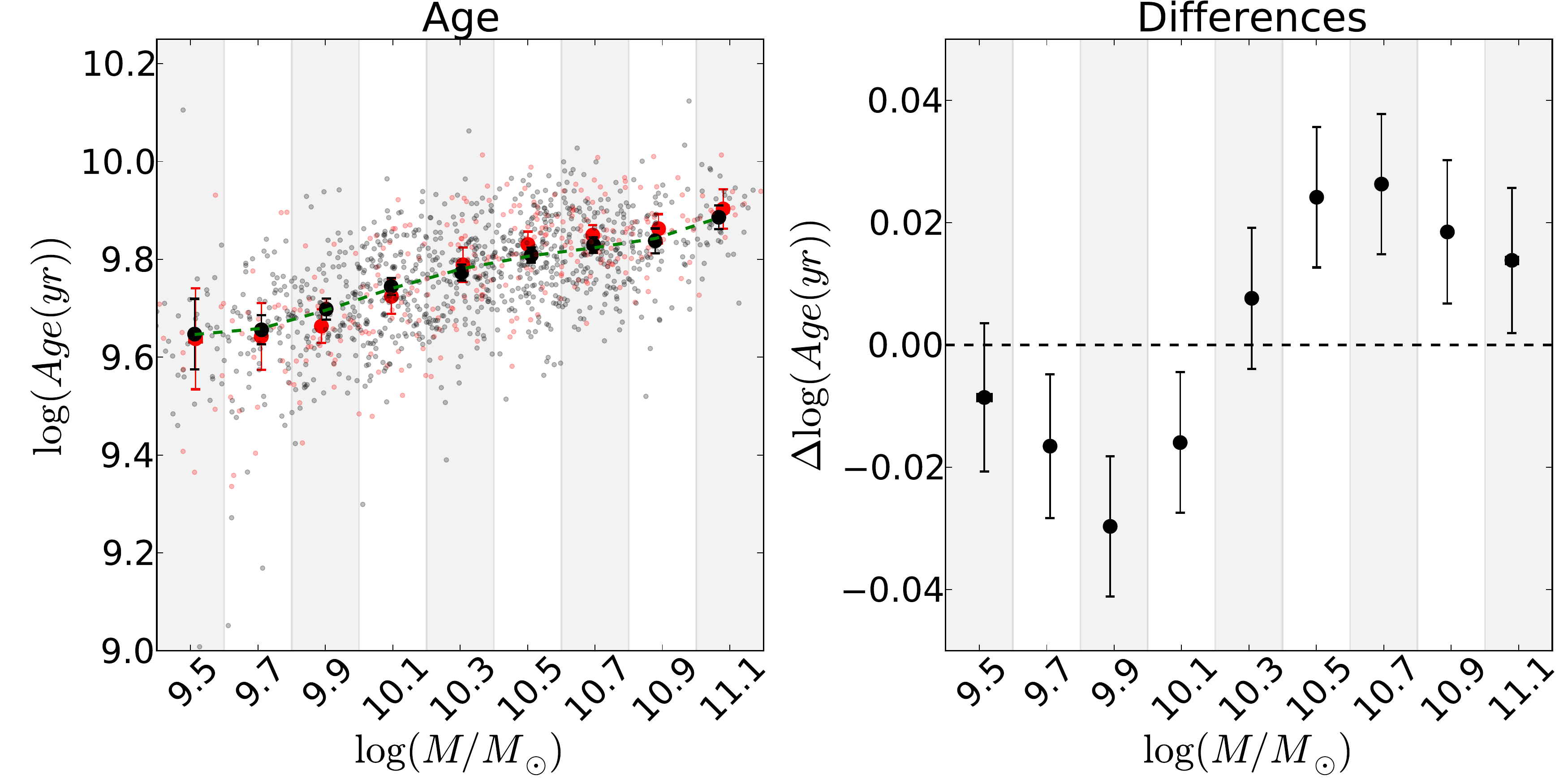}
\caption{(Left panel) Small points represent each individual galaxy (red for barred galaxies, black for unbarred galaxies); large points represent mean ages and metallicities of the galaxies in each mass interval. The green line is a polynomial fit to the unbarred galaxies. The error bars represent the errors in the mean. (Right panel) Differences of barred galaxies from the polynomial. None of the differences were significant (3$\sigma$)}
 \label{figure_massage}
\end{figure}

In Figure \ref{figure_massage} (left panel) we show the known trend of increasing age with mass. In the right panel, we plot the differences in the age between barred and unbarred galaxies. The differences are not significant ({\it P}-values $\geq$ 0.069) so the mean ages of the stellar populations in barred and unbarred galaxies are the same at a given mass. \cite{Coelho12} found that the distribution of the age of high mass ($M_{*}>10^{10.1}M_{\odot}$) barred galaxies is bimodal, with peaks at 4.7 and 10.4 Gyr, whilst the distribution of unbarred galaxies is unimodal. However, the mean age of barred and unbarred galaxies in a mass interval is the same, which is compatible with our results.

\section{Comparison with previous works.}
\label{cap:previous} 

\subsection{Gaseous-phase metallicity}
\label{Previous_gas}

Studies during the 1990s (see \citealt{Dutil99, Considere00} or the review by \citealt{Henry99}) compared the central metallicities of barred and unbarred galaxies finding lower or equal metallicities in the centres of barred galaxies. However, these works relied on small samples, biased towards late-type galaxies. 

The recent work by E11, however, compared a large sample of barred and unbarred galaxies that were selected to cover similar ranges of mass, inclination and redshift. The authors found that, at a given mass, the central metallicity of barred galaxies is significantly larger (by $\sim$0.06~dex) than that of unbarred galaxies, while the SFR of barred galaxies is higher than that of unbarred by $\sim$60 per cent, but only in those galaxies more massive than $10^{10}$M$_{\odot}$. 

We are unable to reproduce this result, despite the fact that we selected our sample in a similar way to them, using the \cite{Nair10} catalogue: for example, the presence of the bar and its characteristics are identical in the two studies. 

There are three main differences between this study and that of E11 and we explore if any (or several) of them are responsible of the different results:
\begin{enumerate}
\item We made our sample of barred and unbarred galaxies equivalent in the distribution of mass, redshift, inclination and morphological type. In E11 the two subsamples were not equivalent in morphological type.
\item We obtain the fluxes from the OSSY data base, whereas E11 used the MPA/JHU data base. Because of this, we perform a different cut in the S/N of the lines, which leads to a different sample selection.
\item The last difference with E11 is that we use R$_{23}$ instead of $[$NII$]$/$[$OII$]$ as a metallicity 
indicator.
\end{enumerate}

{\bf Sample selection:} 
As noted above, the main difference in the sample selection between the present work and that of E11 is that we make our sample equivalent in the distributions of mass, redshift, inclination and morphological type for barred and unbarred galaxies. In E11, the authors also built equivalent distributions in the first 3 parameters, but not in the fourth one.

We compared the distribution of morphological types in the subsamples of barred and unbarred galaxies taken from E11. This comparison shows that in E11 there is an excess of early-type unbarred galaxies relative to barred galaxies. This excess is removed in our sample when we make the distributions of morphologies equivalent. As explained in Section \ref{cap:section_gas_morphology}, the differences in metallicity between barred and unbarred galaxies are larger in early-type than in late-type galaxies. If the sample is biased to early-type galaxies, the differences in the metallicities can be larger.

\begin{figure*}
\includegraphics[width=0.8\textwidth]{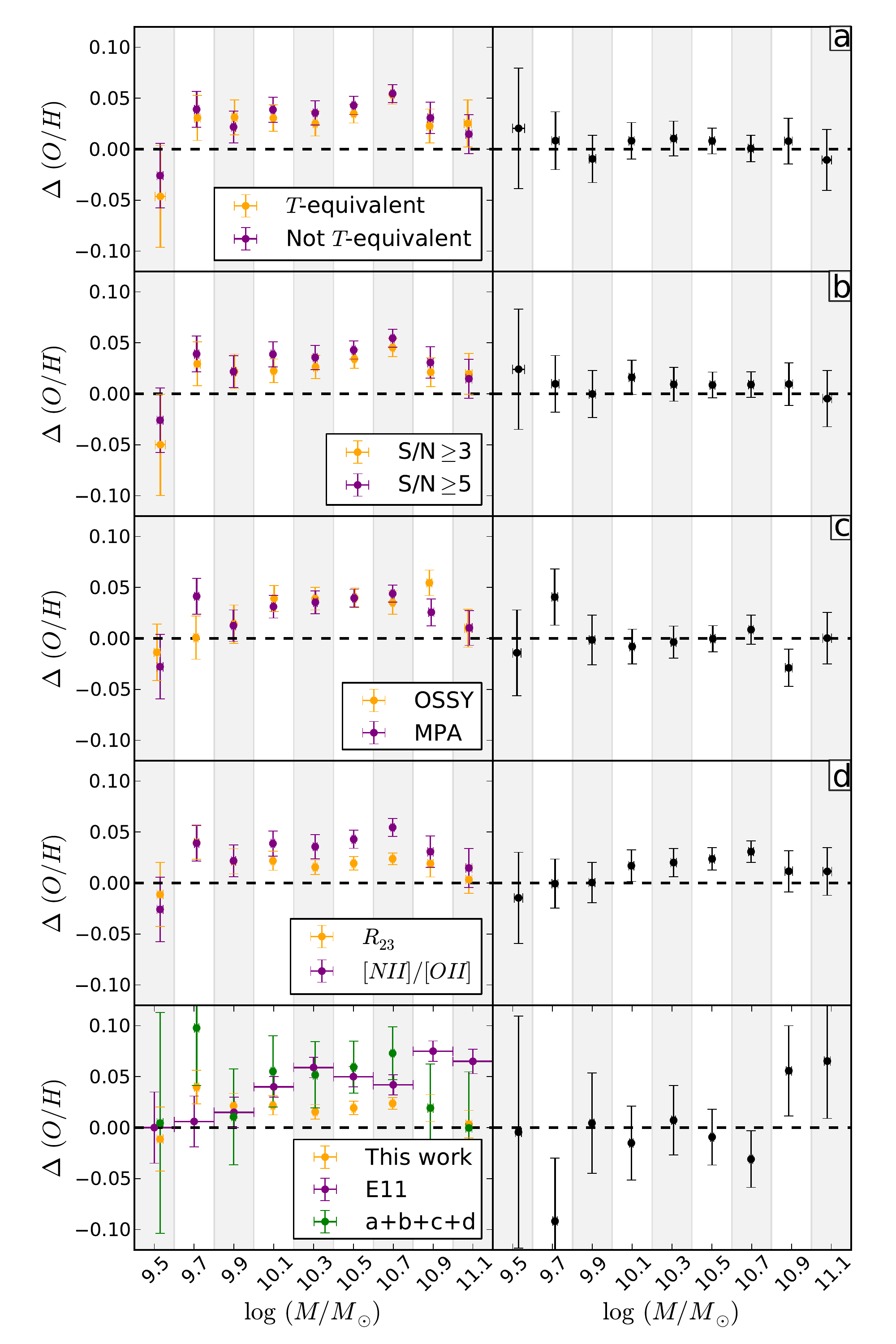}
\caption{(Left panels) The results obtained with the variation of the parameters, from top to bottom fourth panel: equivalence of morphology distributions, S/N cuts, emission-line flux data base and metallicity calibration. The corresponding right panels show the discrepancies found between the results obtained in the left panels. (Bottom left panel) The comparison between E11 (purple) and this work (orange), and this work plus the expected differences (green) arising from different sample selection, S/N cuts, flux data bases and metallicity calibration. The bottom left panel shows the differences between purple and green points. In this case, error bars are compatible withe the discrepancies between works being negligible.}
 \label{figure_differences}
\end{figure*}

In order to study the effects of the sample selection, we made new samples without taking into account the equivalence of the morphology distributions between barred and unbarred galaxies. The first row in Fig. \ref{figure_differences} shows this comparison. The left panel shows the differences found between the subsamples of barred and unbarred galaxies selected with equivalent morphology distributions and the subsamples of barred and unbarred galaxies without using that criterion. The right panel shows the differences between orange and purple points, which correspond to the discrepancies expected between the results using different criteria. As can be seen, the differences between barred and unbarred galaxies are larger in mean ($5\times10^{-3}$~dex) when the morphology distributions are not equivalent. However, this difference is very small and well within the errors.

{\bf Flux measurement:}. There are two effects related with the chosen data base and we are going to study them separately. The first one is the different cut in the emission-liness signal-to-noise ratio.

As we mention above, the signal-to-noise of the emission-liness in the OSSY and MPA/JHU data base are different and related, roughly as 3:5 (see Fig.~\ref{figureSN}). Despite the fact that our cut at S/N$=$3 is similar to that of E11 (which is S/N$=$5), we explore what the result would be if we considered only galaxies with fluxes in the main emission-liness with a signal-to-noise ratio above 5.

The second row in Fig. \ref{figure_differences} shows the differences between the gaseous metallicities of barred and unbarred galaxies with $S/N\geq3$ and $S/N\geq5$ in their emission-liness (left panel) and the discrepancies in the results (right panel). The discrepancies are small, but systematic, and show that using different cuts in the signal-to-noise ratio can affect the results, with larger differences ($9\times10^{-3}$~dex) when using $S/N\geq5$ than when using $S/N\geq3$. However, as in the case above, the differences are very small in themselves and cannot explain the discrepancies found with E11.

The second effect comes from the differences in the flux values themselves, which are different in the OSSY data base (chosen in this work) and in MPA/JHU (used in E11). In Fig.~\ref{figure_fluxes} we showed that [OIII]/[OII], is higher in unbarred galaxies by 0.010~dex. We do not know the origin of these differences, but because this ratio is involved in the metallicity calculation, the discrepancy could lead to different results depending on the data base used.

We calculated the metallicities with MPA/JHU fluxes and compared them with the metallicities obtained using OSSY. The comparison can be seen in the third row of Fig. \ref{figure_differences}. In this case, the choice of data base does not have as much influence on the metallicities (less than $1\times10^{-3}$~dex) as other parameters, and no clear discrepancies are seen in relation to this choice.

{\bf Metallicity calibration:} We calculated the metallicities using the $R_{23}$ calibration, but E11 used a modified method of \cite{Kewley02}, which is explained in \cite{Kewley08}. Basically, this method consists in using [NII]/[OII] instead of $R_{23}$ calibration for $12+\log({\rm O}/{\rm H})>8.5$, that is, for all galaxies in this sample. As this is a required step in our calculation of metallicities, we keep the metallicities obtained with this calibration.

The fourth row in Fig. \ref{figure_differences} shows how the choice of calibration can alter the results. The differences between barred and unbarred galaxies are clearly larger (0.01~dex in mean) when the metallicity is calculated using [NII]/[OII] than when using $R_{23}$. The explanation for this could be that [NII]/[OII] is less sensitive to the ionization parameter than $R_{23}$, and we saw in Fig. \ref{figure_SFR_logq} that the distribution of the ionization parameter is different in barred and unbarred galaxies. However, the differences relative to the metallicity calibration are still a factor of 6 smaller than those found by E11.

Finally, we calculated the estimated total discrepancies between our work and E11 by adding the discrepancies found when varying all the former parameters.

This is shown in the fifth row of Fig. \ref{figure_differences}. The left panel shows the differences between barred and unbarred galaxies that we and E11 found. Also plotted are the differences we found between barred and unbarred galaxies adding the differences we find owing to different methodologies. The right panel shows the differences between green and purple points on the left panel. 

As can be seen, although the differences obtained by modifying each parameter at a time are always smaller than the differences found by E11 in their comparison of barred and unbarred galaxies, the sum of all effects together, namely differences in the morphology distributions, different signal-to-noise cuts and the difference in the calibration index, can explain the differences, except for the two most massive bins.

\subsection{Stellar-phase metallicity}

Regarding the stellar populations, much less work has been carried out. \cite{Moorthy06} and \cite{Perez11} found hints of barred galaxies being more metal-rich than unbarred galaxies. However, their methodology and sample are very different from ours. They analysed long-slit spectra and calculated the metallicity gradients inside the bulge, whereas we used integrated spectra. Furthermore, their sample was much smaller and restricted to early-type galaxies. We cannot say that our results disagree with previous results as we are studying integrated properties and their work is more focused on resolved properties. Moreover, the signal-to-noise in the continuum required for a detailed analysis of the stellar population is very high. 

In this work, taking into account the error bars and the confidence level used, we are only able to find differences in the metallicity $\Delta$[Z/H]$\geq0.022$ dex, which is larger than the differences found in \cite{Perez11} and \cite{Moorthy06}.

We also expanded the morphological range to late-type galaxies. This expansion is in agreement with the work of \cite{Coelho12}, in which they do not find significant differences between barred and unbarred galaxies considering all morphological types. They also found a low-age (4.7 Gyr) barred galaxies population, without an unbarred counterpart. We do not see any differences in the age of barred and unbarred galaxies in our sample, but our analysis is different than that from \cite{Coelho12} as we compare the mean age of galaxies at a given mass, not the distribution, and we remove AGNs from our sample.

\section{Discussion and conclusions}\label{cap:discussion}

In the present study we compared a large sample (1595) of low-inclination disc galaxies with and without bars from SDSS. This allowed us to obtain reliable results and to be very restrictive with the selection criteria.

\subsection{Mass as independent variable}

Throughout the whole paper we take the mass as the reference for comparing barred and unbarred galaxies, and this is the physical magnitude that drives the evolution of galaxies. Measuring the mass is not an easy task, and is always an indirect measure that depends on several other magnitudes and requires a series of assumptions. One of the magnitudes it strongly depends on is the M/L ratio, which also depends on the age and the metallicity of the stars.

We use MILES \citep{Sanchez-Blazquez06, Vazdekis10} for calculating the stellar parameters (see Appendix \ref{Starlight_Steckmap_Lick} for more information), whereas in \cite{Kauffmann03b}, where the masses are taken from, BC03 models \citep{Bruzual03} are used to calculate the stellar mass in the galaxies. The way in which each set of models is built is very different, so the results obtained with each set is different from each other. In this paper we obtained ages and metallicities different from those given in \cite{Nair10}. Moreover, \cite{Kauffmann03b} only build models at solar metallicity, but the M/L ratio in general depends on the metallicity.

The aim of this paper is not to discuss the precision or the errors in the determination of the mass. However, as far as the mass is determinated in the same way for barred and unbarred galaxies and the results obtained using the mass do not change when the luminosity is used instead, we consider that our results are robust no matter what the absolute values of mass, age and metallicity are.

\subsection{Gas phase metallicities}

We calculated the gaseous metallicities of these galaxies using their central spectrum (from SDSS) and emission-line fluxes from the OSSY data base. We compared the metallicities of barred and unbarred galaxies from a number of perspectives. We use $R_{23}$ for the metallicity calibration and the methodology described in \cite{Kewley02}.

We found that, in most cases, the differences in the central metallicities are not significant (at 3$\sigma$ level, $P-value$=0.0027), but barred galaxies are systematically more metal-rich than unbarred galaxies. This is particularly true for early-type galaxies (S0 to Sb) for which the differences tend to decrease with increasing mass, while for late-type galaxies they tend to increase with mass.

We compared our work with previous results. We found discrepancies and a possible explanation for them. We found that some metallicity calibrations are more sensitive to differences in the metallicities than others. The mean differences found using $R_{23}$ calibration were around 0.02~dex, but using [NII]/[OII] as calibration, the metallicity differences increased to 0.04~dex. We fond that discrepancies between studies are caused by a number of factors, such as selection criteria, how the fluxes are calculated, and how the metallicity is obtained. These criteria can change the results in $\sim$0.04~dex, which is a large value compared with the differences in the metallicities.

Our conclusion is that barred galaxies are slightly more metal-rich than unbarred galaxies. To understand the origin of these differences, we compared the SFRs of barred and unbarred galaxies. While the total SFR is the same for barred and unbarred galaxies, barred galaxies have a slightly higher efficiency per unit of (stellar) mass. This can be explained by gas inflows to the central parts of the galaxies and an enhanced star formation. A comparison of the ionization parameter shows that the gas is more ionized at the centres of barred galaxies, which indicates a larger amount of ionizing radiation, which could be caused by a greater number of massive stars. These more massive stars could be responsible for the larger central metallicities in barred galaxies.

\subsection{Stellar phase metallicities}

We carried out the same analysis on the stellar metallicities. In this case we used the spectral synthesis technique through {\tt STECKMAP}. With this technique, the age-metallicity degeneracy can be avoided to a large extent, leading to reliable results compatible with those obtained using Lick/IDS indices (see Appendix \ref{Starlight_Steckmap_Lick} for more information).

We found no significant differences in any case, which is in agreement with previous studies, such as that by \cite{Coelho12}. We also expanded the morphological range, comparing all morphological types from S0 to Sm galaxies, and the size of the sample with this work. The large errors in the stellar metallicities do not allow us to detect differences in [Z/H] below 0.022 dex, which, given the metallicity range, is clearly insufficient.

\subsection{Gaseous versus Stellar parameters}

We compared gaseous and stellar properties for the first time; that is, the results obtained from emission-liness were compared with the results obtained with the spectral fitting. Surprisingly, we found no correlation between them, which means that the evolution of gas and that of stars are not related, at least in the centres of the galaxies.

We compared the metallicities of both gaseous and stellar phases with the mean age of the stellar populations. Again, we did not observe any relation between gas and stars, but we found a correlation between the stellar age and the metallicity. This correlation is what one would expect if the age-metallicity degeneracy were broken.

Finally , we compared the age of the stellar populations with the mass of the galaxy. We obtained a positive correlation between the two parameters, which is as expected. We did not find significant differences between the ages of barred and unbarred galaxies per unit mass.


\section*{Acknowledgments}\label{cap:acknowledgements} 
We have received financial support through the Spanish research project AYA2010-21322-C03-03, from the Spanish Ministerio de Econom\'ia y Competitividad. PSB is supported by the Ministerio de Econom\'ia y Competitividad (MINECO) through its Ram\'on y Cajal program. IP acknowledges the support by the Spanish Ministerio de Econom\'ia y Competitividad (MINECO) (via grants AYA2010-21322-C03-02, AYA2010-21322-C03-03, AYA2007-67625-C02-02 and Consolider-Ingenio CSD2010-00064) and
by the Junta de Andaluc\'ia (FQM-108).

\bibliographystyle{mn2e}
\bibliography{raul}
\bsp
\label{lastpage}


\appendix
\clearpage
\section{Robustness of the results with the redshift and the inclination}\label{Robustness_inclination}
\label{appendix_1}
\subsection{Redshift effects}

\begin{figure*}
\includegraphics[width=0.9\textwidth]{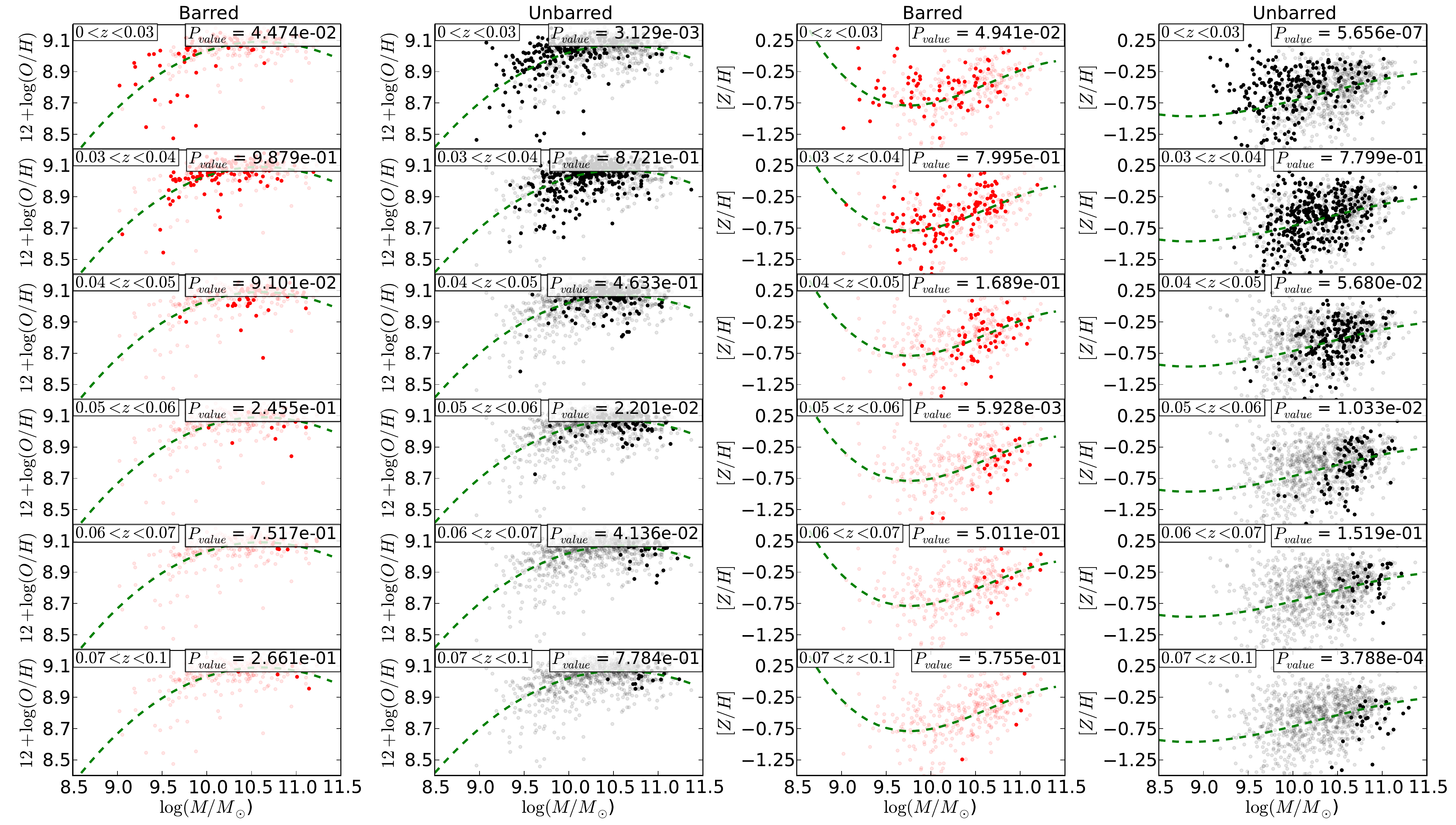}
\caption{Mass-metallicity relations for the galaxies in various redshift bins (given in the top left corner in each plot). Pink/grey points represent barred/unbarred galaxies from the whole sample, while red/black points represent barred/unbarred galaxies in each bin. The green dashed line represents a third-order polynomial fit to pink/grey points. The $P$-value resulting from a KS test comparing the residuals of the (barred or unbarred) galaxies in the bin and the residuals of all (barred or unbarred) galaxies from the fit. If $P$ $\leq 0.027~(3\sigma)$ the subsamples are not statistically representative of the whole sample.}
\label{figure_robustness_redshift}
\end{figure*}

Redshift is not a physical magnitude, but is a parameter to be considered because galaxies at different redshifts can behave in different ways. Figure \ref{figure_robustness_redshift} shows a comparison between the metallicities of barred and unbarred galaxies in different redshift intervals and of barred and the whole subsamples of barred and unbarred galaxies. We compared the residuals (for both gas and stars) of barred and unbarred galaxies in each bin from a polynomial fit (to the whole subsample of barred and unbarred galaxies) with the residuals of the whole corresponding subsample. These residuals are compared using KS tests, whose $P$-values (which are the probability that the differences are random) can be seen in the top right corner of each plot. As can be seen, all $P$-values are larger than 3$\sigma$ ($P$-value$\geq$0.027) which means that the metallicities of barred and of unbarred galaxies have the same behaviour in every redshift interval.

\begin{figure*}
\includegraphics[width=0.9\textwidth]{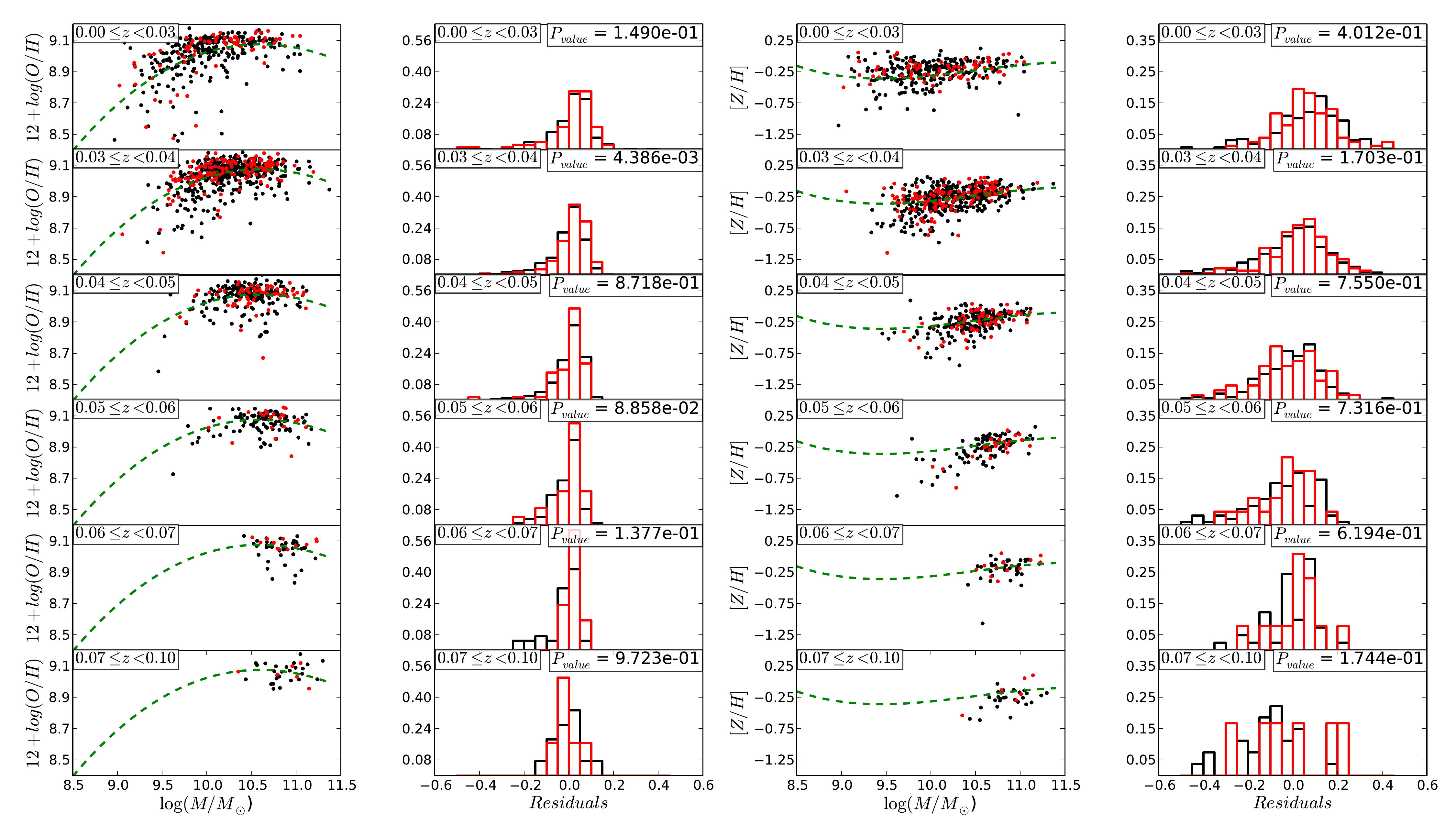}
\caption{Differences in the metallicity between barred and unbarred galaxies in various redshift bins. The first (third) column shows the gaseous (stellar) metallicity versus the mass for barred (red points) and unbarred (black) points. The second (fourth) column shows the residuals from the gaseous (stellar) metallicities to the polynomial fit (green lines). $P$-values are the probability that the differences in the residuals arise from random effects. The confidence level used is $3\sigma$, corresponding to $P$=0.027.}
\label{figure_metal_redshift}
\end{figure*}

We also compared the differences in the metallicity in those bins. We undertook the same analysis as in Sec. \ref{gas_mass_metallicity} in each redshift bin. This can be seen in Fig. \ref{figure_metal_redshift}. We compared the residuals of barred and unbarred galaxies in each redshift bin from a third-order polynomial fit to the whole sample of galaxies. Again, all $P$-values indicate that, at 3$\sigma$, our results are robust under variation of redshift cuts.

The size of the SDSS fibre is fixed (3 arcsec), and may cover different regions of the galaxies, depending on their size, inclination and redshift. In Section \ref{cap:sample_selection} we explained how we constrained our sample, choosing subsamples of barred and unbarred galaxies with equivalent distributions of mass, redshift and inclination, so we are confident that the covering factors have the same distributions for barred and unbarred galaxies.

\subsection{Inclination effects}

Inclination of galaxies can have an effect on the metallicities owing to the disc contribution to the spectrum. Throughout the paper we choose $b/a \geq 0.4$, but here we are going to check the robustness of the results under variation of the inclination criterion. To do that, we follow the same steps as in the previous section.

\begin{figure*}
\includegraphics[width=0.8\textwidth]{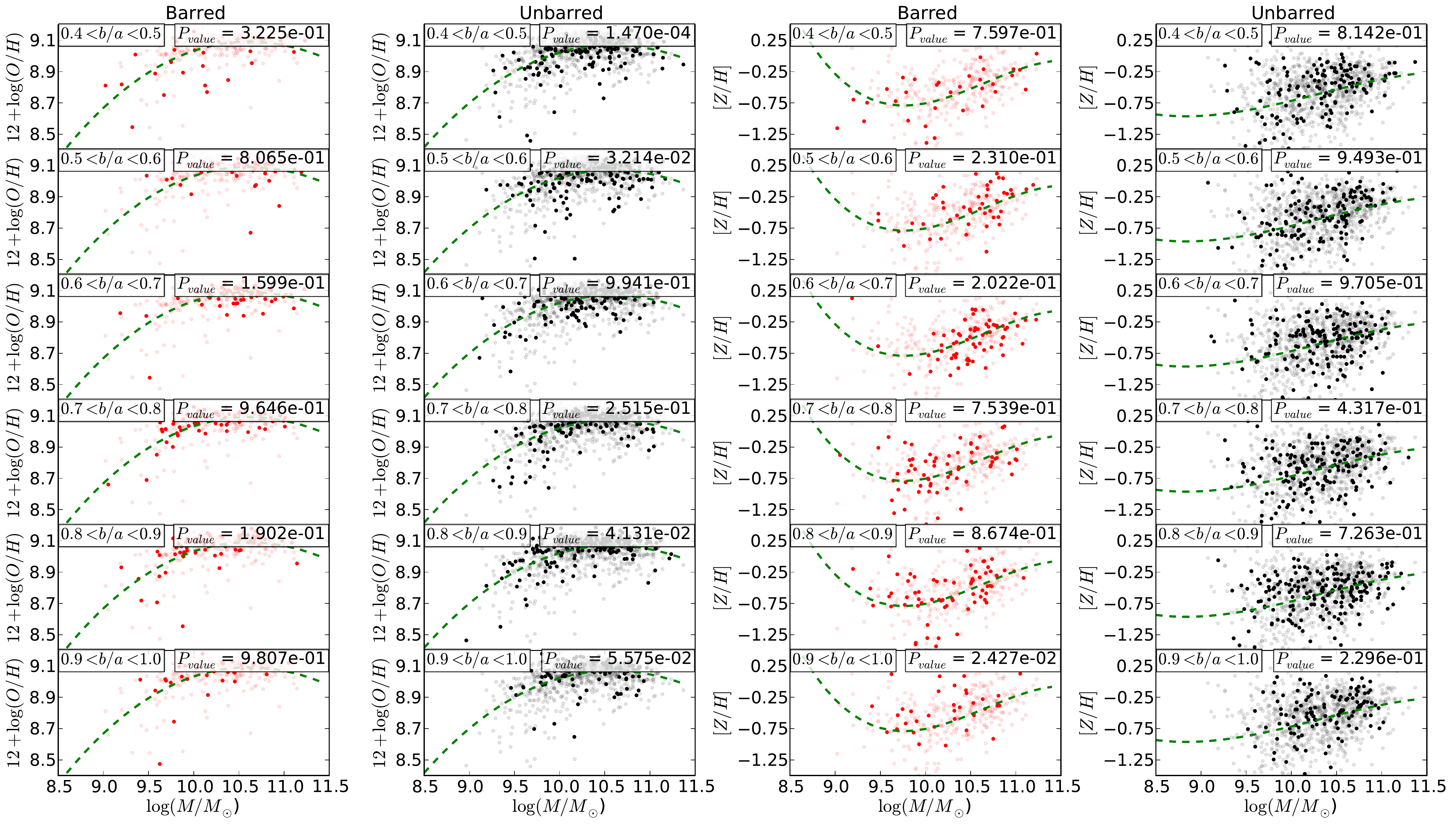}
\caption{As Fig. \ref{figure_robustness_redshift} but using inclination bins instead of redshift bins.}
\label{figure_robustness_b_a}
\end{figure*}

We first compared the mass-metallicity relation of barred and unbarred galaxies in different inclination bins, shown in Fig. \ref{figure_robustness_b_a}. As can be seen, $P$-value$\geq 0.027~(3\sigma)$ in all cases but for unbarred galaxies with $0.4 \leq b/a < 0.5$, which is , which is $1.47\times 10^{-4}$.

\begin{figure*}
\includegraphics[width=0.8\textwidth]{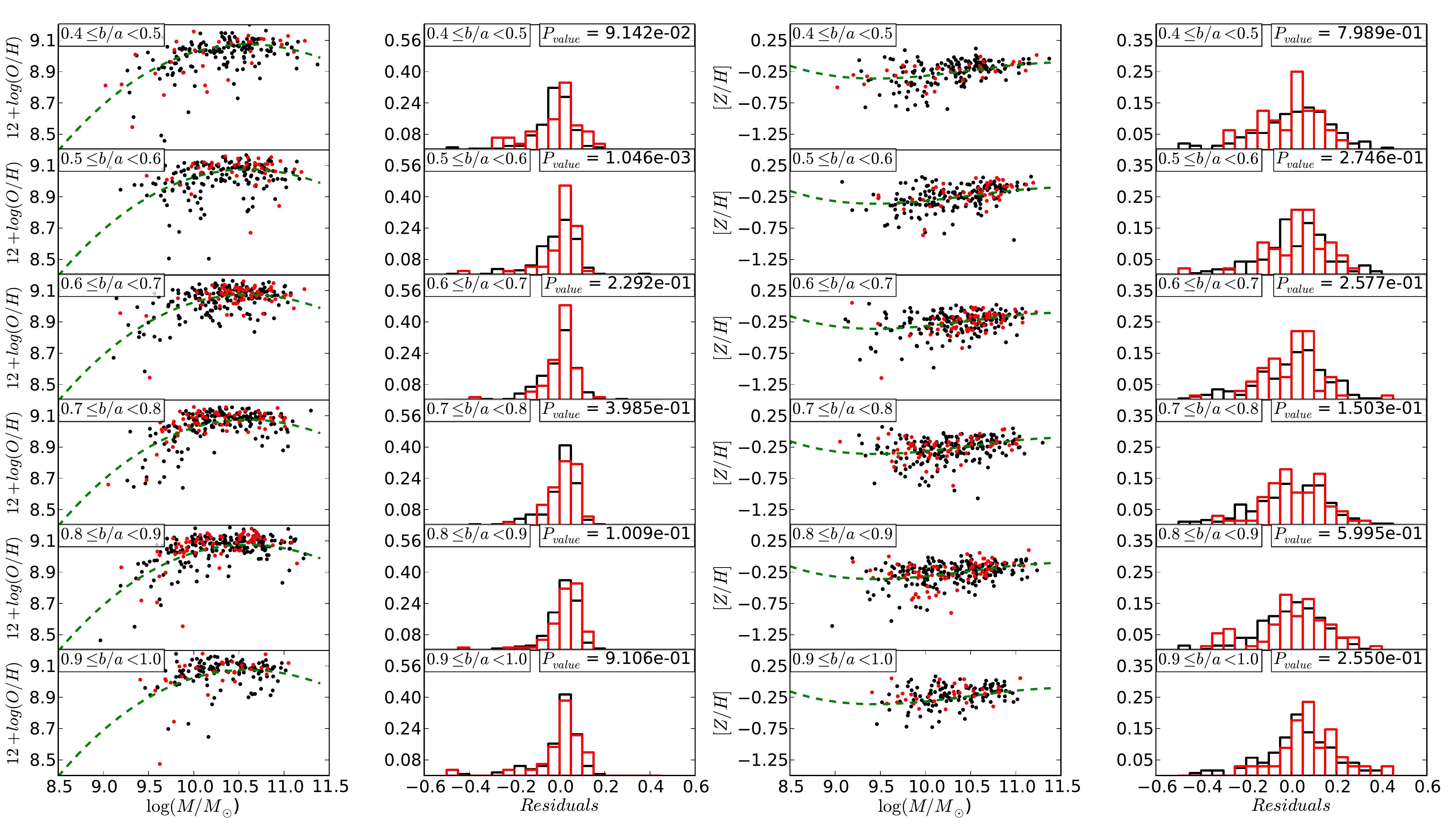}
\caption{Same as Fig. \ref{figure_metal_redshift} but using inclination bins instead of redshift.}
\label{figure_metal_b_a}
\end{figure*}

We compared the differences in the metallicity, in different inclination bins, between barred and unbarred galaxies. This comparison is shown in Fig. \ref{figure_metal_b_a}. It can be seen that the differences in the gaseous-phase metallicities between barred and unbarred galaxies are not significant for the interval $0.4 \leq b/a < 0.5$, but they are for the next one, $0.5 \leq b/a < 0.6$. The differences in the stellar metallicities are not significant in any case. 

We repeated the analysis using only galaxies with $0.5\leq b/a \leq 1.0$ and the results do not vary, so we retained the initial selection criteria.

\section{Comparison between {\tt STECKMAP}, {\tt STARLIGHT} and Lick/IDS indices}\label{Starlight_Steckmap_Lick}

Part of this work is based on the results obtained with {\tt STECKMAP} \citep{Ocvirk06a, Ocvirk06b} a software package that fits the spectrum of a galaxy using stellar models \citep[in our case, we used MILES, ][with a Kroupa universal initial mass function, \citealt{Kroupa01}]{Sanchez-Blazquez06}. With this, we can calculate the mean properties of the underlying stellar populations. We also used another code, {\tt STARLIGHT} \citep{CidFernandes05}, which uses a different approach, for comparison.

A more classical technique is that of Lick/IDS indices \cite{Worthey94}. This technique consists in plotting two indices on a grid of models. The position of the point gives the age and the metallicity of the equivalent stellar population of the galaxy.

In order to study the reliability of the spectral fitting, we compared the results obtained with {\tt STECKMAP}, {\tt STARLIGHT} and Lick/IDS indices. ({\tt STARLIGHT} and indices also need stellar population models, so we also used MILES).

\begin{figure}
\includegraphics[width=0.5\textwidth]{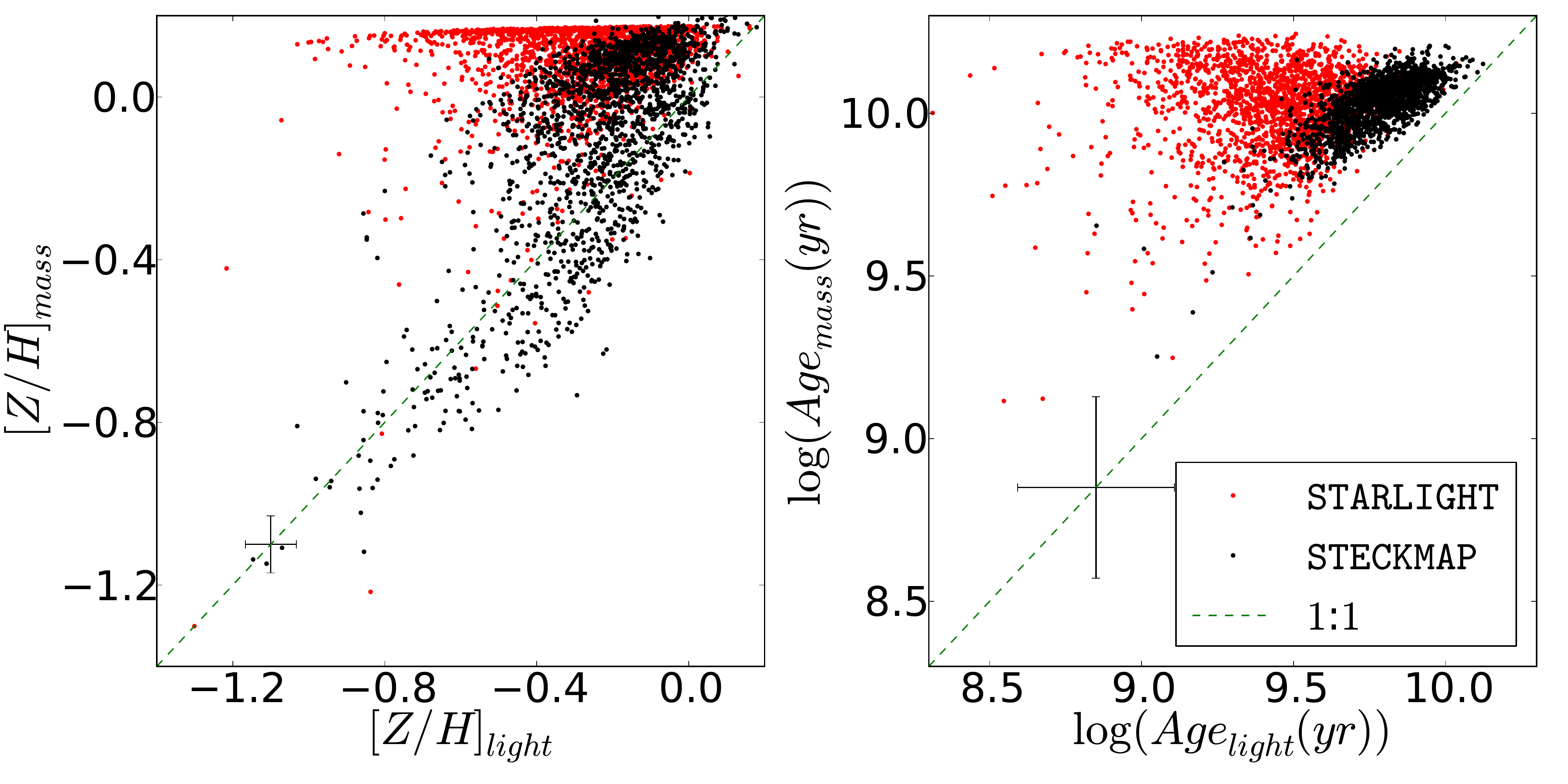}
\caption{Comparison of $L$-weighted versus $M$-weighted values for the full spectral fitting software. The red and black points represent quantities derived with STARLIGHT and STECKMAP respectively. The green dashed line represents the equivalence line. Error bars indicate the typical uncertainties in STECKMAP.}
\label{figure_lightmassvalues}
\end{figure}

Figure \ref{figure_lightmassvalues} shows the mass-weighted values versus luminosity-weighted values for {\tt STARLIGHT} and {\tt STECKMAP}. As can be seen, the results obtained with {\tt STECKMAP} are more consistent, but the scatter in the metallicity is large and there is an offset in the ages. {\tt STARLIGHT}, on the other hand, tends to high mass-weighted metallicities and low luminosity-weighted ages.

\begin{figure}
\includegraphics[width=0.5\textwidth]{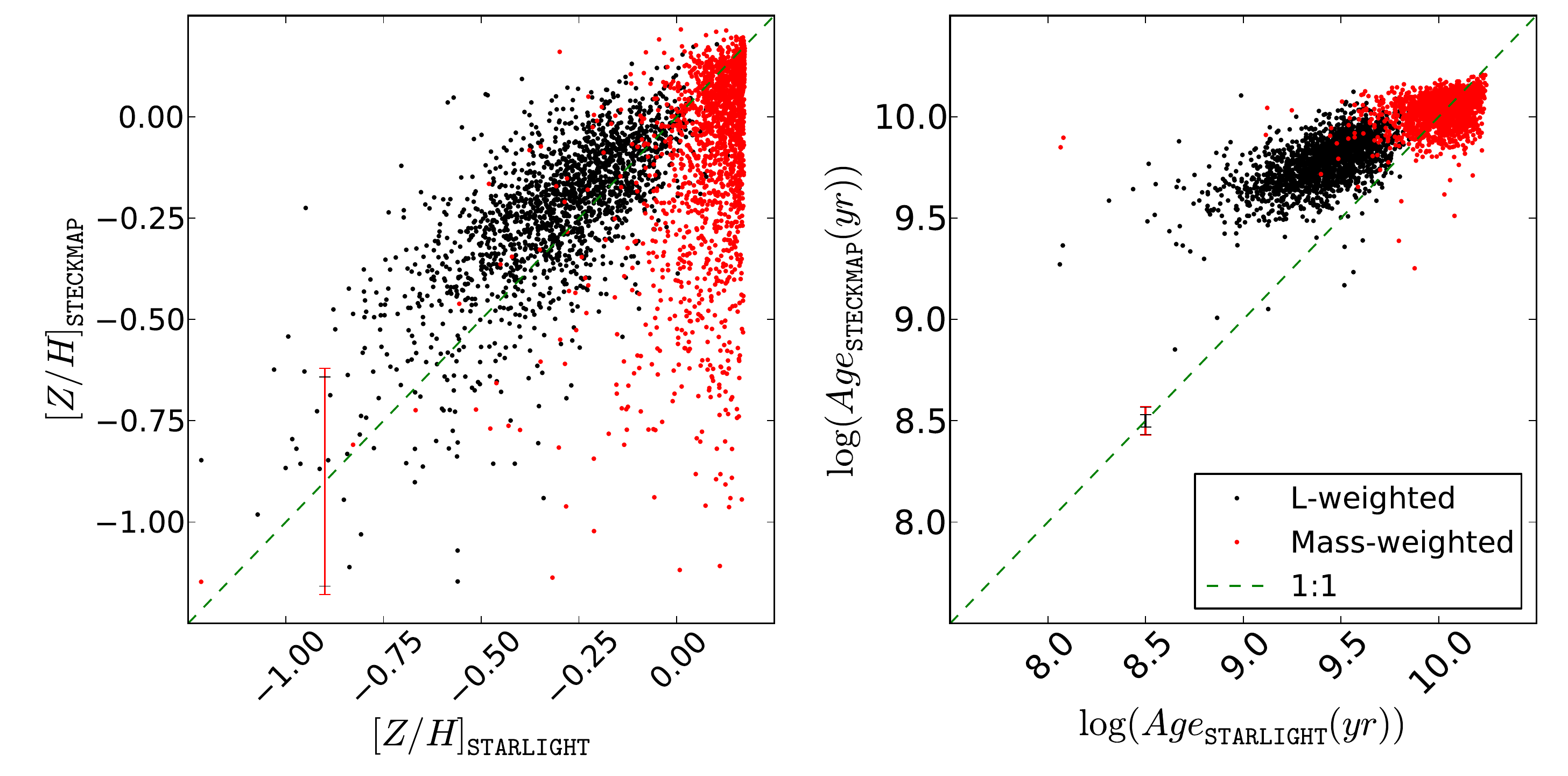}
\caption{Comparison of the results obtained with STARLIGHT and STECKMAP. The red and black points represent mass- and luminosity-weighted values, respectively. Error bars represent the typical uncertainties in the STECKMAP values. }
\label{figure_starlightsteckmap}
\end{figure}

Figure \ref{figure_starlightsteckmap} shows the results obtained with {\tt STECKMAP} and those obtained with {\tt STARLIGHT}. In this case, luminosity-weighted values are in agreement (with some scatter and an offset in the case of the ages), but mass-weighted values are not even well correlated, so we use only luminosity-weighted values.

\begin{figure}
\includegraphics[width=0.5\textwidth]{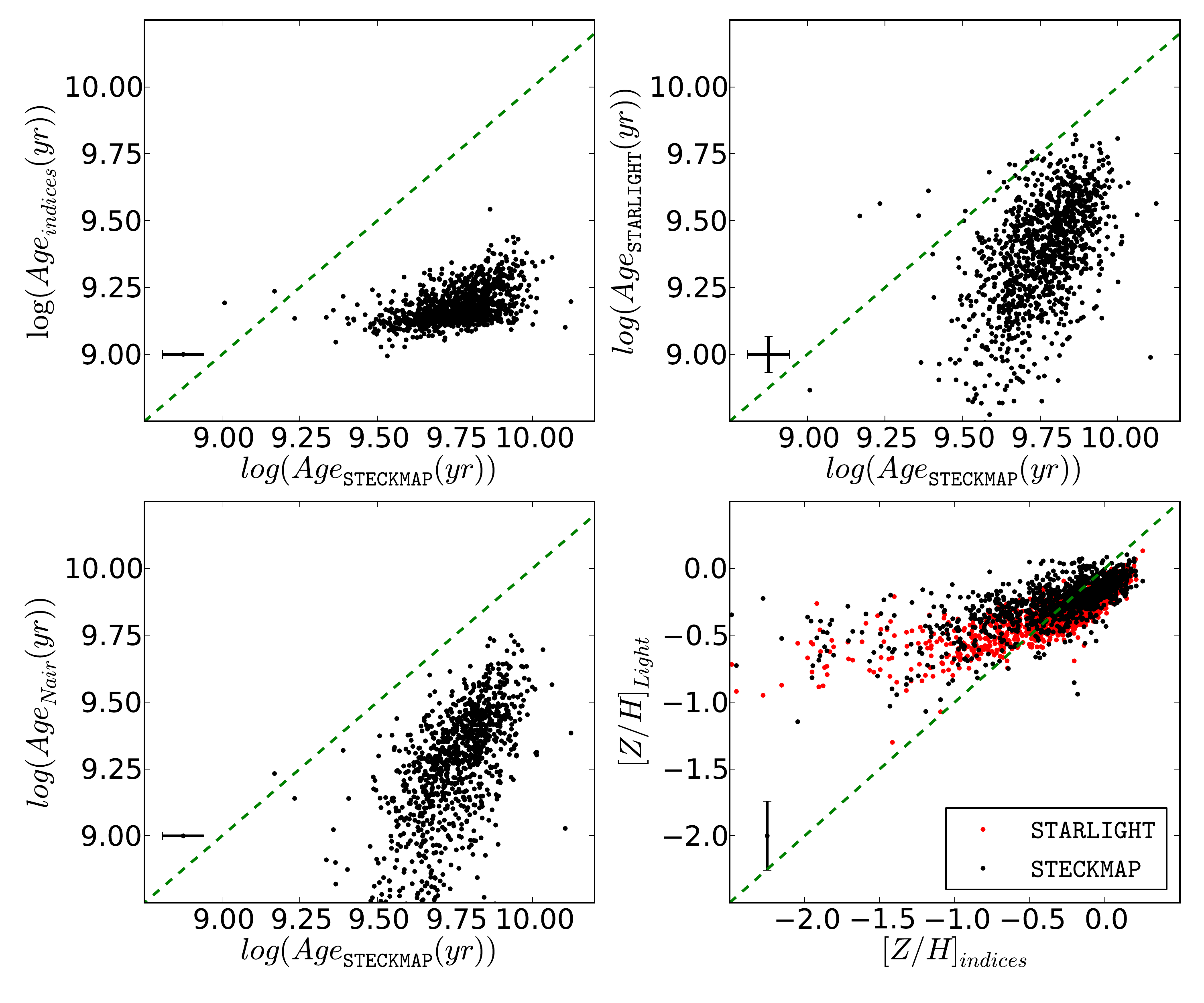}
\caption{Comparison of ages and metallicities obtained through different methods. Top left panel: age obtained with indices versus the age obtained with {\tt STECKMAP}. Top right panel: comparison of the results obtained with {\tt STARLIGHT} and {\tt STECKMAP}.  Bottom left panel: comparison of the ages given in \citet{Nair10} and the ages calculated with {\tt STECKMAP}. Bottom right panel: Comparison of the metallicites calculated with {\tt STECKMAP} (black points) and {\tt STARLIGHT} (red points) against the metallicities calculated through indices.}

\label{figure_indicessynthesis}
\end{figure}

In Fig. \ref{figure_indicessynthesis} we plot the comparison of the results obtained with spectral synthesis (using {\tt STECKMAP}, {\tt STARLIGHT}) and the data given in \cite{Nair10} versus the results obtained with Lick/IDS indices. We compare the ages, although these refer to different things in the two cases. The age obtained with the spectral fitting is the mean age of the stellar populations weighted with the light of the stars, whereas the age obtained with lick/IDS indices is the age of an equivalent SSP. When comparing the metallicities (only for the spectral synthesis methods, because there is no information about the metallicity in the source catalogue), the results are better. We analysed the correlation among the results using the Spearman correlation coefficient (see Tables \ref{table_Spearman_Age} and \ref{Table_Spearman_Z_H} for details). As can be seen there is some correlation among the ages calculated in different ways (while the absolute values are very different), so it is possible to compare ages obtained from spectral fitting with ages from indices. The correlation among metallicities is stronger, and even the cloud of points is very close to the 1:1 line for high values of the metallicity.

\begin{table}
\begin{tabular}{ c | c | c | c | c}
 & Indices & {\tt STARLIGHT} & {\tt STECKMAP} & Nair\\
\hline
Indices & --- & 0.66 & 0.45 & 0.42  \\

 &  & ($10^{-147}$) & ($10^{-57}$) & ($10^{-51}$)   \\

 {\tt STARLIGHT} & 0.66 & --- &  0.54 & 0.55   \\
 & ($10^{-147}$) &  &  ($10^{-90}$) &($10^{-91}$)   \\

{\tt STECKMAP} & 0.45 &  0.54  & --- & 0.43    \\
& ($10^{-57}$) &  ($10^{-90}$) &  & ($10^{-52}$)    \\

Nair & 0.42 &  0.55 & 0.43 & ---  \\
 & ($10^{-51}$) &  ($10^{-91}$) &($10^{-52}$)   & \\

\end{tabular}
\caption{Spearman correlation coefficients for the ages appearing in this paper \citep[Calculated with lick/IDS indices, STARLIGHT, STECKMAP and taken from][]{Nair10}. The closer the coefficient is to 1, the harder the correlation is. Below each value, in brackets the P-Value can be found, which is the probability of finding the Spearman coefficient if the subsamples are not correlated.}
\label{table_Spearman_Age}
\end{table}

\begin{table}
\begin{center}
\begin{tabular}{ c | c | c | c | c}
  & Indices & {\tt STARLIGHT} & {\tt STECKMAP} \\
\hline
Indices & --- & 0.86 & 0.72 \\
 &  &  ($0.0$) & ($10^{-182}$) \\

 {\tt STARLIGHT} & 0.86 & --- &  0.69 \\
  & ($0.0$) &  &   ($10^{-159}$) \\

{\tt STECKMAP} & 0.72 &  0.69 & ---  \\
 & ($10^{-182}$) & ($10^{-159}$) &   \\

\end{tabular}
\end{center}
\caption{Same as Table \ref{table_Spearman_Age} but using [Z/H] instead of the age.}

\label{Table_Spearman_Z_H}
\end{table}

The spectral fitting technique is a powerful tool for obtaining the stellar population parameters. Not only are its results correlated with the results obtained with indices, but it is also able to estimate the contribution of different underlying stellar populations. Furthermore, as shown in Fig. \ref{figure_metalages}, it breaks the age-metallicity degeneracy, which is a difficulty when using indices.

In light of the above, we decided to use the spectral fitting technique to obtain our results. The decision to use {\tt STECKMAP} was based on Fig. \ref{figure_lightmassvalues}, which shows that the results obtained with {\tt STECKMAP} are more consistent between luminosity- and mass- weighted values.


\end{document}